%% file: main.tex
\documentclass[preprint,12pt]{elsarticle}
\input{my_include.tex}

\journal{Physica D}

\begin{document}

\begin{frontmatter}

\title{Control of diffusion-driven pattern formation behind a wave of competency}

\author[inst1,corresponding]{Yue Liu}

\affiliation[inst1]{organization={Mathematical Institute, University of Oxford},%Department and Organization
            %addressline={Woodstock Road}, 
            city={Oxford},
            %postcode={OX2 6GG}, 
            %state={State One},
            country={UK}}
            
\affiliation[corresponding]{Corresponding author: yue.liu@maths.ox.ac.uk}

\author[inst1]{Philip K. Maini}
\author[inst1]{Ruth E. Baker}

\begin{abstract}
In certain biological contexts, such as the plumage patterns of birds and stripes on certain species of fishes, pattern formation takes place behind a so-called ``wave of competency". 
Currently, the effects of a wave of competency on the patterning outcome is not well-understood.
In this study, we use Turing's diffusion-driven instability model to study pattern formation behind a wave of competency, under a range of wave speeds. Numerical simulations show that in one spatial dimension a slower wave speed drives a sequence of peak splittings in the pattern, whereas a higher wave speed leads to peak insertions. In two spatial dimensions, we observe stripes that are either perpendicular or parallel to the moving boundary under slow or fast wave speeds, respectively. 
We argue that there is a correspondence between the one- and two-dimensional phenomena, and \revision{hypothesize that (as others have)} pattern formation behind a wave of competency can account for the pattern organization observed in many biological systems.
\vspace{\baselineskip}
\end{abstract}

\begin{highlights}
\item Effects of a wave of competency on diffusion-driven pattern formation.
\item Determination of directionality and alignment of stripe patterns.
\item Stability of patterns under the effect of additive structural noise.
\item Comparison of wave of competency with growing domain mechanisms.
\end{highlights}

\begin{keyword}
Turing pattern \sep diffusion-driven instability \sep pattern formation
\sep wave of competency \sep partial differential equations
\MSC 92C15 \sep 37N25 \sep 35B30
\end{keyword}

\end{frontmatter}

\section{Introduction}\label{sec:intro}

The diffusion-driven instability (DDI) mechanism proposed by Turing \citep{turing1952ChemicalBasisMorphogenesis} has been used as a canonical model for pattern formation in mathematical biology \citep{murray2001mathematical}. Originally, the model was proposed as a mechanism for cell fate specification, but it has since been adopted to describe a wide range of biological phenomena \citep{kealy2012NonlinearStabilityAnalysis,cartwright2002LabyrinthineTuringPattern,watanabe2015PigmentPatterningFish}. In this framework, cell fate is hypothesised to be determined by underlying spatially pre-patterned chemical cues.
Turing termed these chemicals morphogens, and proposed that different rates of diffusion of the morphogens in the system destabilize a spatially homogeneous steady state, leading to spatially inhomogeneous chemical profiles (patterns). The basic applications of DDI models are on fixed domains, with kinetic parameters constant throughout the entire domain (see e.g. \citep[Ch.~2.1]{murray2001mathematical}). 

There have been many extensions to the basic DDI model proposed in order to obtain a richer, and more robust range of patterning behaviours, reflecting those observed in biology \citep{krause2021IntroductionRecentProgress}. 
One such extension is domain growth. %For the cell fate models, domain expansion can be interpreted as the growth of the underlying tissue where the patterning take place. 
The most basic form of growth, where the underlying tissue grows uniformly, results in dilution effects on the morphogens, which can lead to interesting behaviours not found on a fixed domain, as reported by
\citep{barrass2006ModeTransitionsModel,madzvamuse2007VelocityinducedNumericalSolutions,crampin1999ReactionDiffusionGrowing,crampin2002PatternFormationReactiondiffusion,vangorder2021TuringConditionsPattern}, among others. However, in many cases, tissue growth happens on a much slower timescale compared to pattern formation, therefore the effects of domain growth are ignored as they are not biologically relevant.

In contrast, it has been hypothesized that, in certain cases, a related mechanism called a ``wave of competency" may serve as a more biologically relevant way to generate interesting patterning behaviours.
In this setting, the overall domain size is fixed, but it is divided into two subdomains by a propagating front representing the so-called wave of competency. Patterning can take place only in the subdomain that lies behind the propagating wave of competency, and this subdomain expands as the wave advances. %, which can take place on a similar timescale as pattern formation. 
In the context of patterning via a DDI, we assume that the model parameters are such that the sub-system behind the wave of competency is capable of displaying a DDI, while that ahead of the wave is not.
The goal of this paper is to understand the behaviour of the patterning sub-system under different speeds of propagation of the wave of competency.

One possible biological interpretation for the wave of competency is as follows. The overall fixed domain represents the entirety of the underlying tissue. The subdomain in front of the wave represents immature tissue that cannot support patterning until it matures further, i.e. becomes ``competent";  and the subdomain behind the wave represents mature tissue, forming the effective domain of patterning.
Experimental support for the existence of a wave of competency comes from the plumage patterning of birds, for example
\citet{bailleul2019SymmetryBreakingEmbryonic} argued that early mesodermal development leads to the formation of feather fields (the patterning subdomains), which expand on a similar time scale to that of pattern formation.

In an earlier study, \citet{jung1998LocalInhibitoryAction} identified possible molecular candidates encoding the wave of competency for feather formation, and \citet{jiang1999SelforganizationPeriodicPatterns} provided further experimental evidence.
\citet{mou2011CrypticPatterningAvian} examined a similar mechanism that restricts feather formation to one side of a front.
The study \citep{bailleul2019SymmetryBreakingEmbryonic} found that while a DDI model on a fixed domain could reproduce the final pattern of follicles in birds, it was not able to reproduce the correct sequence of emergence of the feather follicles, nor the consistent orientation of the stripe-shaped transient structures known as feather tracts, which exist prior to the formation of follicles. \revision{This is because with a biologically realistic initial condition, which is usually taken to be a homogeneous steady state plus a noisy perturbation, patterns emerge simultaneously throughout the domain.}
Hence, they proposed a model which combined chemotaxis and DDI mechanisms by coupling the dynamics of two morphogens with the evolution of cell density. In their model, the cell population undergoes logistic growth and diffusion, in addition to responding chemotactically to one of the morphogens. The morphogen concentrations are governed by a DDI model, which is coupled to  the cell density model via the reaction terms.
If the cell density is assumed to be a constant parameter, then the DDI system possesses a stable spatially uniform steady state, which becomes unstable when the cell density becomes sufficiently high.
This means that the region of high cell density forms the patterning subdomain, which is initially small, but expands outwards under the effects of chemotaxis and diffusion.
\revision{This model exhibits sequential pattern formation behaviour, in agreement with observations of feather formation.}

\revision{
Here, we seek to better understand this type of patterning scenario by isolating the effect of increases in the size of the patterning subdomain on the dynamics of a model capable of exhibiting a DDI.
The resulting model can be viewed as a classical DDI model with a minimal modification that represents a wave of competency, which is analogous to cell density in the feather model developed by \citep{bailleul2019SymmetryBreakingEmbryonic}. In essence, the wave of competency takes the form of a moving boundary which divides the domain into two subdomains. Within each subdomain, the system behaves as a classical reaction-diffusion system, but the dynamics differ between the subdomains, with a DDI only possible for the subdomain that lies behind the moving boundary. We are interested to understand whether such a modification could enable the model to reproduce sequential pattern formation.}

\begin{figure}
    \centering
    \captionsetup[subfigure]{labelformat=empty}
    \begin{subfigure}[h]{0.3\textwidth}
        \centering
        \captionsetup{justification=centering}
        \includegraphics[width=\textwidth]{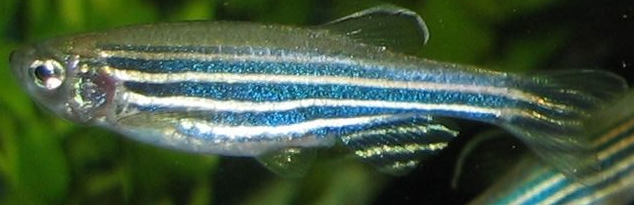}
        \caption{Zebrafish (\textit{Danio rerio})}
    \end{subfigure}
    \begin{subfigure}[h]{0.3\textwidth}
        \centering
        \captionsetup{justification=centering}
        \includegraphics[width=\textwidth]{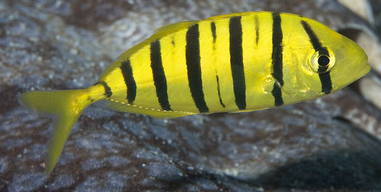}
        \caption{Golden Trevally \\(\textit{Gnathanodon speciosus})}
    \end{subfigure}
    \begin{subfigure}[h]{0.3\textwidth}
        \centering
        \captionsetup{justification=centering}
        \includegraphics[width=\textwidth]{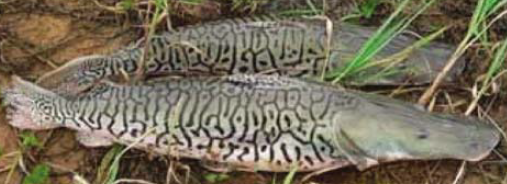}
        \caption{Catfish (\textit{Pseudoplatystoma tigrinum})}
    \end{subfigure}
    \caption{Example of fishes with different pigmentation stripe orientations. 
    The Zebrafish has horizontal stripes \citep{zebrafishpic}, whereas the Golden Trevally has vertical stripes (\citep{trevallypic}, image under CC-BY license), and a certain species of catfish has a more complex pattern where the stripes are not strictly aligned (figure reprinted with permission from \citep{barrio2009ModelingSkinPattern}).
    }
    \label{fig:fish_pics}
\end{figure}

In the classical fixed domain setting, typical patterns arising from DDI in two spatial dimensions take the form of spots, stripes, or labyrinths. 
We will  focus on understanding the effect of a wave of competency on stripe or labyrinthine patterns, since stripe patterns can exhibit greater variation through their directionality and alignment than typically observed in spot patterns.
Futther, In the natural world, some stripe patterns exhibit certain preferential directions, and stripes may or may not be aligned throughout the domain.
To illustrate stripe directionality and alignment, we use pigmentation patterns on fishes as an example. 
As shown in Fig.~\ref{fig:fish_pics}, the stripes on a zebrafish are always aligned head-to-tail, along the antero-posterior axis, whereas some other fish species, such as the banded trevally and certain types of angelfish, have their stripes aligned vertically instead, along the dorso-ventral axis. However, certain catfish species can develop stripe patterns that are not strictly aligned, but nonetheless display a preference for directionality.
\revision{It is still an open question as to the precise mechanisms driving the alignment of these patterns. 
For zebrafish, \citet{kondo1995ReactionDiffusionWave} suggest that the stripes are the result of a Turing mechanism, however \citep{owen2020QuantitativeModellingApproach,volkening2015ModellingStripeFormation,frohnhofer2013IridophoresTheirInteractions} suggest an alternative mechanism based upon cell-cell interactions. Here, we will use Turing's DDI model to explore stripe directionality and alignment in general.}

In contrast to the aligned stripes and consistent directionality seen in nature, the typical patterns produced by a DDI model under spatial perturbations to a homogeneous steady state, however,  tend to be disordered labyrinthine patterns without clear alignment or directionality (examples can be seen in Fig.~\ref{fig:sim2d_noisyinit}, last column).
This raises the question of how, and whether, aligned stripes and consistent directionality might be achieved.
In this paper, we suggest that expansion of the patterning subdomain behind a wave of competency, where different wave speeds select different preferred directions, can effectively control pattern alignment and directionality.

There have been several other proposed mechanisms for pattern alignment. \citet{shoji2002DirectionalityStripesFormed} showed that anisotropic diffusion can account for directionality, while \citet{hiscock2015OrientationTuringlikePatterns} showed that a gradient in the reaction parameters, or anisotropic growth, can do the same, and \citet{nakamasu2009InteractionsZebrafishPigment} were able to produce aligned stripes with a specific initial condition.
\citet{page2005ComplexPatternFormation} considered parameters that vary across space in more complicated ways, and showed that the DDI model, is, in this case, able to produce a larger variety of patterns.
\revision{Finally, \citet{frohnhofer2013IridophoresTheirInteractions} suggested prepatterning as another mechanism.}
We will compare the patterns arising from implementations of these different mechanisms with the one we are proposing in the discussion.

Although the main motivation behind the study of DDI models comes from biology, it is very challenging to experiment directly with proposed DDI systems in this context due to the complexity of living tissues. Therefore, chemical systems have been used to experimentally validate DDI models, since their dynamics are better understood, and more amenable to manipulation.
In \citet{konow2019TuringPatternsRadially}, the authors experimentally observed and simulated the CDIMA (chlorine dioxide-iodine-malonic acid) reaction, which is capable of displaying DDI, on a patterning subdomain bounded by a circular boundary. This boundary is then made to expand radially outward, serving as an analogue of a wave of competency. In both the chemical experiment and numerical simulations of the corresponding mathematical model, Konow et al found that the stripes produced behind such a wave of competency tend to be concentric circles aligned parallel to the moving boundary if the expansion speed is fast, and radial stripes aligned perpendicular to the boundary if the expansion speed is slow.

In this paper, we seek to determine whether this effect is specific to the setting in \citep{konow2019TuringPatternsRadially}, or if it is generalizable to a wider range of scenarios. 
We will mainly consider Schnakenberg kinetics \citep{schnakenberg1979SimpleChemicalReaction}, a simple model that allows insight into the observed dynamics, as well as the CDIMA kinetics.
\revision{We will show that for the CDIMA model, we can replicate the behaviours observed in \citep{konow2019TuringPatternsRadially}. Considering both models allows us to draw conclusions on whether the observed phenomena are particular to the CDIMA model, or may be observed more generally in models that can exhibit a DDI. By comparing the behaviours of these models in one dimension versus two dimensions, we are also able to provide insight into why variations in the wave speed affect stripe directionality and alignment. We also define a speed for the natural propagation of patterns, and illustrate its significance in determining patterning outcome. This has, to our knowledge, not been considered in previous studies.}

In Section~\ref{sec:models}, we describe the models to be used, and provide details of how the wave of competency and the expanding subdomain are implemented. In Section~\ref{sec:numerics}, we use a series of numerical simulations to explore the behaviour of the model, and to identify and distinguish the cases where different wave speeds lead to qualitatively distinct patterns. Finally, we summarize our findings in Section~\ref{sec:discuss}.

\section{Models and methods}\label{sec:models}

The basic two-morphogen DDI model in two spatial dimensions, with an imposed wave of competency, can be written as the following system of partial differential equations (PDEs),
\begin{subequations}
\begin{align}
\pd{u}{t} &= D_u \nabla^2 u +f(u,v,W), \quad (x,y) \in \Omega, t>0,\\
\pd{v}{t} &= D_v \nabla^2 v +g(u,v,W), \quad (x,y) \in \Omega, t>0,\\
W &= W(x,y,t), \\
\pd{u}{n} &= \pd{v}{n} = 0 ,\quad (x,y) \in \partial\Omega , \nonumber \\
u(x,y,0) &= u_0(x,y), \quad v(x,y,0) = v_0(x,y), \nonumber
\end{align}\label{sys:generic_rd}%
\end{subequations}
where $u(x,y,t)$ and $v(x,y,t)$ are the concentrations of the morphogens U and V, respectively, $D_u$ and $D_v$ are their respective diffusion coefficients, and $f$ and $g$ the reaction functions, which differ for the two models we consider. \revision{We use the function $W(x,y,t)$ to encode the effects of a wave of competency.} We impose zero-flux boundary conditions, since in most applications the morphogen cannot cross the boundary of the domain.
The model is said to display a DDI (also known as a Turing instability) if it possesses a spatially homogeneous steady state that is stable in the absence of diffusion, but can be destabilized in the presence of diffusion by a spatially non-homogeneous perturbation \citep{turing1952ChemicalBasisMorphogenesis}. 
A necessary condition for a DDI is that the diffusion coefficients must be different, and, without loss of generality, we take $D_u \ll D_v$.

\begin{figure}[ht]
    \centering
    \includegraphics[width=0.5\textwidth]{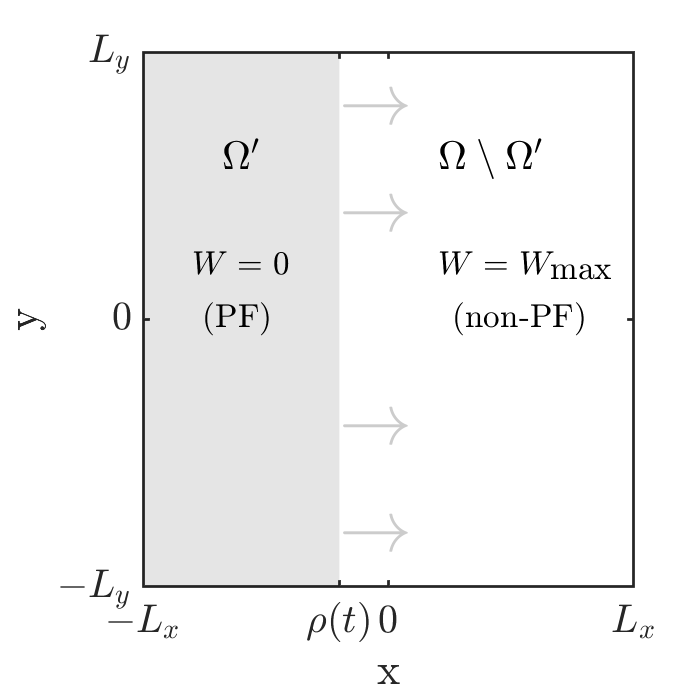}
    \caption{Illustration of the effective patterning domain, $\Omega'$, which is bounded by the wave of competency at $x=\rho(t)$ that propagates towards the right with constant speed $\alpha$. (PF -- pattern forming)}
    \label{fig:domain}
\end{figure}

\revision{We model the wave of competency as a piecewise constant function $W$, such that $W=0$ on the patterning-competent subdomain $\Omega'(t)$, and $W=\Wmax$ on the patterning-incompetent subdomain $\Omega \setminus \Omega'(t)$. The wavefront represents the boundary separating the two subdomains.}
The values for $\Wmax$ and the other model parameters are chosen to ensure that the system is capable of undergoing a DDI in $\Omega'(t)$, but not in $\Omega \setminus \Omega'(t)$. The domain and subdomains are illustrated in Fig.~\ref{fig:domain}.

In \citep{konow2019TuringPatternsRadially}, $\Omega'$ was chosen to be a circular region with the radius increasing linearly with time, at constant speed $\alpha$.
Naturally, we expect the system to be attracted to its homogeneous steady state in $\Omega \setminus \Omega'$ and Turing patterns to form inside $\Omega'$, with the patterns  expanding outwards in synchrony with the expansion of the region $\Omega'$.
Both chemical experiments and numerical simulations carried out by \citep{konow2019TuringPatternsRadially} observed this expected behaviour.
The interesting observation is that the type of pattern that forms depends on the wave speed, $\alpha$, which determines the rate of expansion of $\Omega'$.
For the CDIMA chemical reaction system (see later in this section) on a circular expanding subdomain, the authors mainly observed the formation of concentric ring patterns at higher wave speeds, and stripes perpendicular to the circular front at slower wave speeds, with a mixture of the two behaviours for intermediate values of $\alpha$. The same behaviours are observed in numerical simulations of a DDI model for this reaction. %These observations agreed with numerical simulations of the model.
A reproduction of the numerical simulation results can be found in Fig.~\ref{fig:2d_circular}.

The impact of the wave speed, $\alpha$, upon the pattern structure is the motivation of our study. We hypothesize that the relation between the rate of domain expansion and the rate of pattern formation is the key to distinguishing between the different behaviours.

\revision{It is known that a curved domain boundary can impact pattern formation \citep{barrio1999TwodimensionalNumericalStudy,varea1997ConfinedTuringPatterns,bunow1980PatternFormationReactiondiffusion}. Intuitively, zero-flux boundary conditions force any stripes near the boundary to be either perpendicular or parallel to it, therefore a curved boundary is likely to result in curved stripes, which will impact pattern formation in the interior.}
Thus, in order to remove any potential boundary curvature effects, in this work we will simulate the DDI model described in Eq.~\eqref{sys:generic_rd} on a rectangular domain. 
We take
\begin{equation}
\Omega=[-L_x,L_x] \times [-L_y,L_y], \quad  \Omega'(t)=[-L_x,\rho(t)] \times [-L_y,L_y], \quad  \rho(t)=\min(-L_x+\alpha t, L_x), \label{eqn:domain}
\end{equation}
and recall that
\[W(x,y,t) = \left\{
\begin{aligned}
&0 &&\text{\ if \ } (x,y) \in \Omega'(t)  ,\\
&\Wmax &&\text{\ otherwise} .
\end{aligned}
\right.\]
This choice of domain is illustrated in Fig.~\ref{fig:domain}.
Mathematically, the function $W$, together with Eq.~\eqref{sys:generic_rd}, forms a non-autonomous reaction-diffusion system on a fixed domain.

% \revision{It might help to understand the model equivalently as two autonomous reaction-diffusion system coupled via boundary conditions, as follows.
% \begin{align*}
% (u,v)&=\begin{cases}
% (u_1,v_1), \quad (x,y) \in \Omega', t>0, \\ (u_2,v_2), \quad (x,y) \in  \Omega\setminus\Omega', t>0,
% \end{cases} \\
% \pd{u_1}{t} &= D_u \nabla^2 u_1 +f(u_1,v_1,0), \quad\qquad\  (x,y) \in \Omega', t>0,\\
% \pd{v_1}{t} &= D_v \nabla^2 v_1 +g(u_1,v_1,0), \quad\qquad\  (x,y) \in \Omega', t>0,\\
% \pd{u_2}{t} &= D_u \nabla^2 u_2 +f(u_2,v_2,\Wmax), \ \quad (x,y) \in \Omega\setminus\Omega', t>0,\\
% \pd{v_2}{t} &= D_v \nabla^2 v_2 +g(u_2,v_2,\Wmax), \ \ \quad (x,y) \in \Omega\setminus\Omega', t>0,
% \end{align*}
% and with zero-flux boundary conditions on $\partial \Omega$. Furthermore,  across any boundary between $\Omega'$ and $\Omega\setminus\Omega'$, since there is no physical barrier, and mass must be conserved across the boundary, both $u$ and $v$, and their first spatial derivatives must be continuous. It would be simpler to numerically implement Eq.~\eqref{sys:generic_rd}, so we will refer to that formulation throughout the rest of the paper.
% }

In this paper we consider two specific sets of reaction kinetics: those for the Schnakenberg model, and those that describe the CDIMA reaction, in order to explore whether the effect of the wave of competency depends on the form of the reaction kinetics, or the phase of the patterns. 
For the Schnakenberg model, the profiles for $u$ and $v$ are out-of-phase, whereas for the CDIMA model they are in-phase.

\paragraph{Schnakenberg model}
The model we will primarily focus on is the Schnakenberg model, which was proposed by \citet{schnakenberg1979SimpleChemicalReaction} for a hypothetical reaction system, as a special case of the Gierer--Meinhardt model \citep{gierer1972TheoryBiologicalPattern}. It consists of very simple kinetics, and it is one of the most well-studied DDI models \citep{iron2004StabilityAnalysisTuring}. The non-dimensional reaction terms can be written as
\begin{subequations}
\begin{align}
f(u,v) &= a-u+u^2 v + W ,\\
g(u,v) &= b-u^2 v .
\end{align}\label{sys:schnackenberg}
\end{subequations}
Using linear stability analysis \revision{(see \ref{apx:lsa})} and simulations on two-dimensional fixed domains as a guide, we take as default parameters:
\begin{equation}
D_u=1,\quad D_v=20, \quad a=0.05, \quad b \in \{1.4, 1.6\},\quad L=100,\quad \Wmax=1 .
\label{eqn:schnackenberg_params}
\end{equation}
In this model, $u$ and $v$ represent the concentrations of some abstract morphogens, and we take $W$ to represent an increase in the production rate of $u$. The system possesses the following unique homogeneous steady state when $W$ is treated as a constant,
\begin{equation}
u_* = a+b+W, \quad v_* =b/(a+b+W)^2,\label{eqn:schnackenberg_hss}
\end{equation}
which cannot exhibit a DDI when $W=\Wmax=1$, but can display a DDI when $W=0$.

\revision{We consider two values for $b$, while all other parameters are kept fixed.} On a fixed domain without a wave of competency, with all other parameters taking the values in Eq.~\eqref{eqn:schnackenberg_params},  selecting $b=1.4$ tends to give rise to patterns that consist of a mixture of spots and stripes on a fixed square  two-dimensional domain, while selecting $b=1.6$ mostly gives rise to patterns that consist of stripes (see Fig.~\ref{fig:sim2d_noisyinit}, right-most column).
The conditions which favour spots over stripes or \textit{vice versa} have been investigated on a fixed two-dimensional domain in \citep{ermentrout1991StripesSpotsNonlineara,nagorcka1992StripesSpotsPrepatterns}. Here we are interested to see if imposing a wave of competency can change this behaviour, \revision{and whether the wave of competency interacts differently with spots and stripes}.

\paragraph{CDIMA model}
We also consider the kinetics derived for a model of the CDIMA (chlorine dioxide–iodine–malonic acid) chemical reaction taken from \citet{konow2019TuringPatternsRadially}, which is a modified version of the model by \citet{lengyel1991ModelingTuringStructures}. The non-dimensional form of the model is
\begin{subequations}
\begin{align}
f(u,v) &= a-u-\frac{4uv}{1+u^2}-W ,\\
g(u,v) &= \sigma b \lb u - \frac{uv}{1+u^2}+W \rb ,
\end{align}\label{sys:cdima}%
\end{subequations}
with default parameter values taken from \citep{konow2019TuringPatternsRadially}:
\begin{equation}
D_u=1, \quad D_v = \sigma, \quad a=12, \quad b=0.31, \quad d=1, \quad \sigma=50, \quad L_x=L_y=100, \quad \Wmax=1.5 .
\label{eqn:cdima_params}
\end{equation}
Physically, $u$ and $v$ represent the concentrations of two of the chemical species in the CDIMA reaction, and $W$ represents the amount of illumination applied onto the reactor. This model assumes that the reactant U turns into V at a constant rate, and U and V are consumed together in a reaction that is inhibited by an abundance of U.
When $W<a/5$ is treated as a constant, this system possesses a unique positive homogeneous steady state at
\begin{equation}
u_* = \frac{a-5W}{5}, \quad v_*=\frac{(u_* + W)(1+u_*^2)}{u_*}, \label{eqn:cdima_hss}
\end{equation}
which cannot exhibit a DDI when $W=0$, but can undergo DDI when $W=\Wmax=1.5$.

This model is able to capture the patterning behaviours of the CDIMA chemical system, qualitatively matching the orientation and alignment of the stripes \citep{konow2019TuringPatternsRadially}. 
Since the illumination can be readily controlled, as opposed to the physical size of the reactor, manipulating the illumination provides an easy way of observing DDI behind a wave of competency.

\paragraph{Artificial noise}
Throughout this work, we will add noise to the models throughout their simulation, at each time step. The reasons for this are as follows.

Firstly, there is intrinsic noise present in almost all biological systems, therefore adding noise enables us to better reflect reality. 
Secondly, there have long been criticisms that in many situations, the DDI model is very sensitive to noise and changes in the initial conditions, and so the patterns it produces are unreliable \citep{bard1974HowWellDoes}.
By adding noise to model simulations, we can ensure that our observations and conclusions are robust to noise, and  applicable to biologically-realistic situations.

Finally, it is known that for the type of models we are analysing, the numerical scheme chosen for their simulation can have an impact on both the transient behaviour and the final pattern \citep{madzvamuse2007VelocityinducedNumericalSolutions}, with lower-order methods prone to produce aliasing artifacts \citep{ruuth1995ImplicitexplicitMethodsReactiondiffusion}. In fact, as we will see in Section~\ref{sec:sim2d}, the choice of mesh can also have an impact on the qualitative behaviour.
However, with the addition of sufficient noise, the effects of numerical noise can be \revision{subsumed as part of the overall noise}.

In the next section, we will present and compare results first from simulations without, then with, on-going noise. 
%Any noise we add to the initial condition  will be specified in each case.
For noisy simulations, we add noise to $u$ at each grid point and each time step, independently and identically distributed as $N(0,\mu)$, with $\mu=0.01$.
\revision{These are generated with Matlab's \textit{normrnd} function. All solutions presented are produced with the Mersenne Twister algorithm with seed set to $0$, which is the default setting.}
We found that this magnitude of noise is sufficiently strong to break symmetries and provide new behaviours compared to the noiseless case, and sufficiently weak so that it does not overwhelm the system, i.e. solutions known to be stable on a fixed domain persist in the presence of this noise. 
See \ref{apx:high_noise} for details.

\section{Numerical simulations}\label{sec:numerics}
In this section, we carry out numerical simulations of the models outlined in Section~\ref{sec:models}. We use a standard finite difference scheme for the spatial discretization, where we put equally spaced mesh points at $x_n = -L_x + n \Delta x$,  with $\Delta x=0.5$, and similarly for the $y$ direction in two dimensions. For the time stepping, we use an implicit-explicit scheme (IMEX) \citep{ruuth1995ImplicitexplicitMethodsReactiondiffusion,ascher1995ImplicitExplicitMethodsTimeDependent,canuto1988SpectralMethodsFluid}, where we implement the diffusion term using the implicit Crank–Nicolson method, and use the explicit Euler method for the nonlinear reaction term, with $\Delta t = 0.01$. 
\revision{The Crank--Nicolson scheme has been shown to be stable even in the presence of noise \citep{kamrani2010RoleCoefficientsGeneral,zouraris2018CrankNicolsonFinite}.}
We have verified that all described behaviour still occurs for a much finer time-stepping with $\Delta t=0.00125$ or a finer grid size with $\Delta x=0.25$, and that the numerical scheme indeed converges as $\Delta t$ is decreased.
Model simulations are implemented in Matlab, with the \revision{code, as well as the animations corresponding to the solutions, provided in the supplementary materials and on Github at \url{github.com/liuyue002/turing_expanding_domain}.}

In Section~\ref{sec:sim1d}, we begin with simulating the models on a one-dimensional domain to identify and understand the basic behaviours of the model and guide our choice for the range of wave speeds, and then in Section~\ref{sec:sim2d} we move to two-dimensional domains in order to explore the full range of possibilities for pattern behaviours. Finally, in Section~\ref{sec:sim2d_narrow} we use a very narrow two-dimensional domain to closely examine the structure of the stripes in the patterns observed in the earlier simulations. The dynamics on this narrow two-dimensional domain are effectively one-dimensional, which allows us to explain the differences between the distinct patterns observed under varying speeds of propagation of the wave of competency in two dimensions, by relating them to behaviours in one dimension.

\subsection{Simulations in one spatial dimension}\label{sec:sim1d}

Although the focus of this study is on two-dimensional pattern structures, we begin in one dimension, where the pattern and the transient behaviours are simpler to visualize and understand.
We simulate the PDE system~\eqref{sys:generic_rd} on the one-dimensional domain $\Omega=[-L,L]$. 
Analogous to the two-dimensional case, the patterning subdomain is defined as
$\Omega'(t) = [-L,\rho(t)],$
where $\rho(t)$ is given in Eq.~\eqref{eqn:domain}.
For comparison, we also run simulations in the classical (fixed domain) set up without an expanding patterning subdomain, so that $\Omega' = \Omega$ and $W \equiv 0$.

For the noiseless case, we initialize the system at its homogeneous steady state $u=u_*, \, v=v_*$ (Eq.~\eqref{eqn:schnackenberg_hss}), with a perturbation to $u$ at the left-hand boundary, created by setting $u(-L,0)=2u_*$. 
For simulations with an expanding subdomain, we found that for the range of $\alpha$ we tested, this initial condition produces the same behaviour as for the simulations when a noisy perturbation is added to the initial condition across the whole domain $\Omega$. This is because at the beginning of the simulation, most of the domain is not capable of supporting a pattern, so the noise far from the left-hand boundary simply decays away and has no further impact.
The simulation is run up to 100 time units after the patterning subdomain, $\Omega'$, has expanded to cover the entire domain, $\Omega$, which we found, by visual inspection, is sufficient time for the pattern to evolve to its steady state in both models.

\begin{figure}[H]
    \centering
    \includegraphics[width=0.5\textwidth]{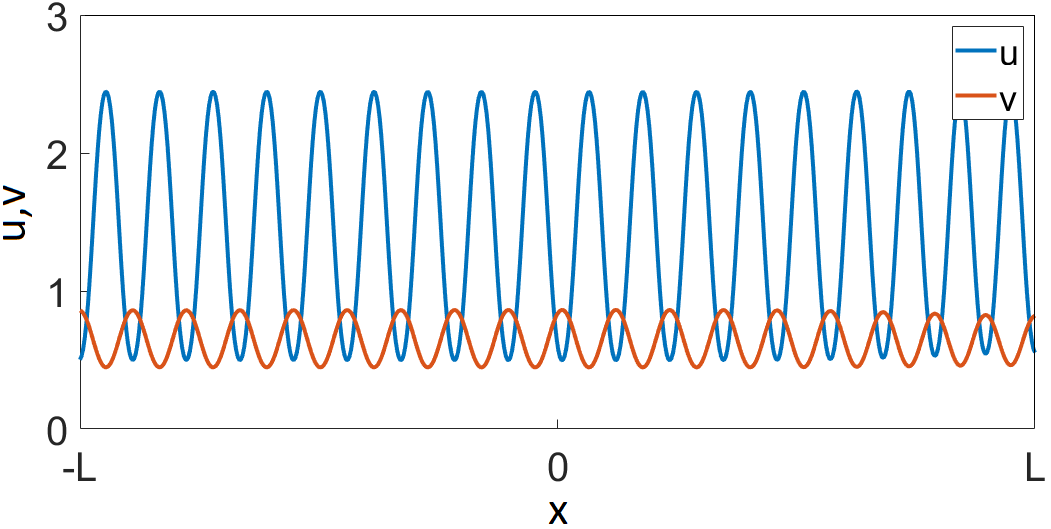}
    \caption{A typical multi-spike steady state solution of the Schnakenberg model (Eq.~\eqref{sys:schnackenberg}) in one dimension. Notice that $u$ and $v$ are exactly out-of-phase, so we will only consider the profile of $u$ henceforth. This simulation is produced with parameters from Eq.~\eqref{eqn:schnackenberg_params} with $b=1.4$ on a fixed domain (i.e. $\Omega'=\Omega$,  $W\equiv 0$) without on-going noise. The development of the system over time leading up to this pattern is illustrated in Fig.~\ref{fig:sim1d_kymo}(d).}
    \label{fig:typical1d}
\end{figure}

In one spatial dimension, \revision{in the regime where $D_u/D_v \ll 1$}, the only possible patterns in both the Schnakenberg and CDIMA models are multi-spike solutions which consist of a series of $N$ equally-spaced, tall, localized peaks, known as spikes (Fig.~\ref{fig:typical1d}).
The stability properties of multi-spike solutions for the Schnakenberg model have been well-studied with asymptotic analysis \citep{iron2004StabilityAnalysisTuring}. Notably, for a given domain length, there is a range of integers $N$ for which the $N$--spike solutions are stable. 

\begin{figure}[H]
    \centering
    \includegraphics[width=1\textwidth]{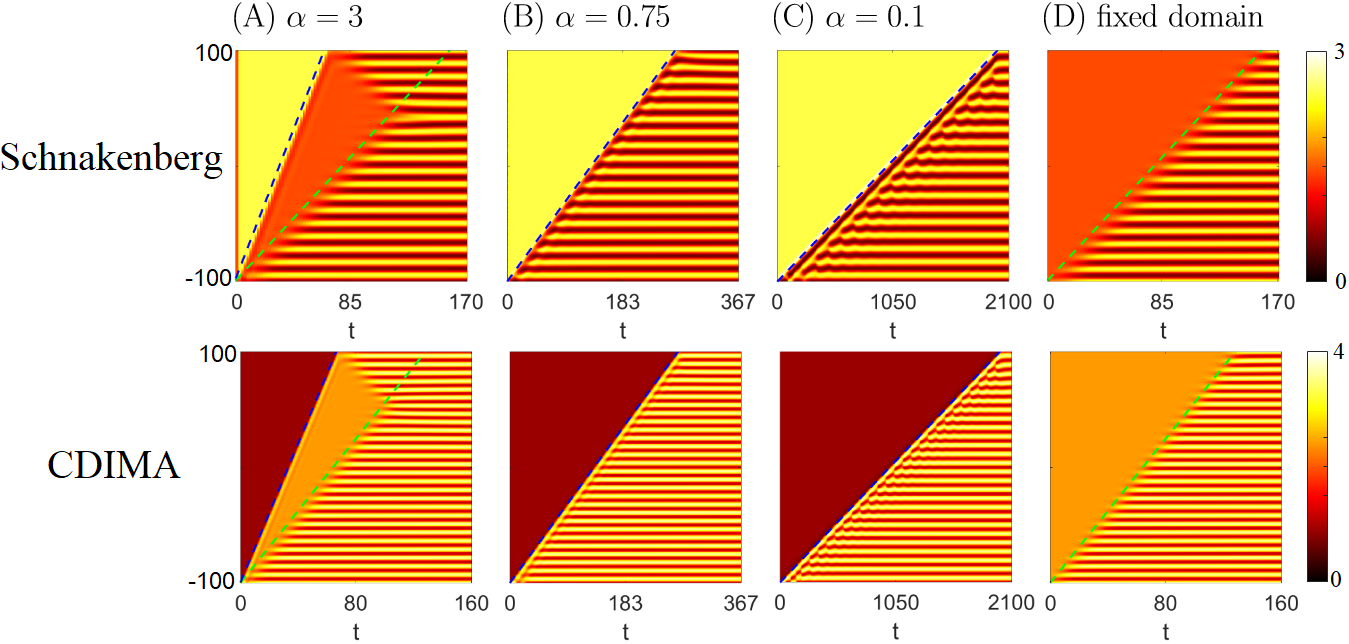}
    \caption{Kymograph representation of $u$ in the solution of the Schnakenberg model (Eq.~\eqref{sys:schnackenberg}) and the CDIMA model (Eq.~\eqref{sys:cdima}) in one spatial dimension, with parameters from Eq.~\eqref{eqn:schnackenberg_params} and $b=1.4$, in the absence of on-going noise. 
    %In this diagram, $t$ is on the horizontal axis and $x$ is on the vertical axis. 
    Each yellow horizontal stripe corresponds to a spike, the red stripes correspond to the valleys between them, and regions of solid colours represent places where the system is locally spatially homogeneous.
    %From left to right: (a) $\alpha=2$; (b) $\alpha=0.75$; (c) $\alpha=0.1$; and (d) fixed domain, as in Fig.~\ref{fig:typical1d}. 
    The system is initialized at the homogeneous steady state with a fixed perturbation at the left-hand boundary, as detailed in the text. Notice that there is a clear divide between the patterning subdomain $\Omega'(t)$ on the lower right (orange with yellow/red stripes), and the non-patterning part of the domain $\Omega \setminus \Omega'(t)$ on the upper left (solid yellow for Schnakenberg, dark red for CDIMA), with the dark blue dashed line representing the wave of competency at \revision{$x=\rho(t)=-L+\alpha t$ separating them. 
    The green dashed line in (A,D) is $x=-L+\nu t$, where the natural speed of pattern propagation is $\nu \approx 1.2428$ for the Schnakenberg model, and $\nu \approx 1.5564$ for the CDIMA model}, see text for details.
    In all four cases, the final pattern is a multi-spike pattern similar to that in Fig.~\ref{fig:typical1d}.
    %The animations corresponding to these simulations are provided at \url{imgur.com/a/pRUZOkn}.
    }
    \label{fig:sim1d_kymo}
\end{figure}

This means that the final pattern, regardless of the wave speed, $\alpha$, will be a multi-spike solution. Indeed, we found that in one dimension, with expanding  patterning subdomains, the final patterns for different wave speeds $\alpha$ are all multi-spike patterns, which are qualitatively similar to the pattern shown in Fig.~\ref{fig:typical1d}, except with possibly different numbers of spikes.
Despite the fact that the form of pattern does not depend on $\alpha$, we will show that the transient behaviour changes as $\alpha$ varies. This transient behaviour will be relevant when we relate the one-dimensional behaviours to two-dimensional behaviours in Section~\ref{sec:sim2d_narrow}, and it is also interesting from a biological perspective since the transient distribution of the morphogens may have consequences on developmental processes. 
We will first discuss the transient behaviours for the Schnakenberg model, shown in Fig.~\ref{fig:sim1d_kymo}, and then note the differences between these and those exhibited by the CDIMA model. \revision{In one dimension, the behaviour of the Schnakenberg model is qualitatively very similar when $b=1.4$ and $b=1.6$, therefore for brevity we only present the results for $b=1.4$.}

First, let us consider the behaviour of the system on a fixed domain without a wave of competency, that is, with $\Omega'=\Omega$ (Fig.~\ref{fig:sim1d_kymo}(D)), to provide a comparison to later simulations.
\revision{
This is simply a standard, classical reaction-diffusion model exhibiting a DDI.
Here the spikes naturally form in sequence, one after another, at a steady rate, hence the pattern propagates away from the initial perturbation at a finite speed. We thus refer to this speed, $\nu$, as the speed of natural pattern propagation as it measures the speed at which an initial perturbation to a homogeneous steady state spreads. This is an intrinsic property of the PDE system and its parameter values, independent of the wave of competency which will be imposed later.
We estimate $\nu$ from the fixed domain simulation as follows.
We use $\tau$, the first time at which $|u(L,\tau)-u_*| > 0.5$, as a measure of the amount of time it takes for the pattern to cover the entire domain. We find $\tau = 160.93$, and with a domain length of $2L=200$, we thus estimate  $\nu=2L/\tau \approx 1.2428$. This steady speed of spike propagation is visualized by the green dashed line in Fig.~\ref{fig:sim1d_kymo}(D).}

Next, we implement a wave of competency advancing at a constant speed $\alpha$, located at $\rho(t)=-L+\alpha t$.
We will consider a range of values of $\alpha$ relative to the natural pattern propagation speed, $\nu$.
First we consider the case where the wave of competency moves more quickly than the natural speed of propagation,  $\alpha > \nu$, as in Fig.~\ref{fig:sim1d_kymo}(A).  
In this case, the subdomain $\Omega'$ expands so rapidly that spike formation, which propagates at speed $\nu$, lags behind the wave of competency, leaving a part of $\Omega'$ (which has width $(\alpha-\nu)t$) just behind the moving boundary momentarily devoid of pattern.
This region corresponds to the small region of solid orange in the plot that sits in between the green and blue dashed lines.
In this case, we see that the moving boundary has no meaningful impact upon the formation of the pattern.
Consequently, we observe that the rate and manner of spike formation, represented by the time of emergence and the shape of the stripes in Fig.~\ref{fig:sim1d_kymo}(A), is nearly identical to what we observe in Fig.~\ref{fig:sim1d_kymo}(D), except near the right-hand boundary $(x=100)$. Here we see that when the front reaches the boundary at $t=100$, 
the higher level of $u$ in the vanishing region $\Omega\setminus\Omega'$ acts as a perturbation to the homogeneous steady state, triggering the formation of a sequence of spikes from right to left.

Next, we consider a moderately high wave speed, \revision{ $0 < \alpha \lesssim \nu$ (denoting a value of $\alpha$ that is below, but close to, $\nu$)}, as in Fig.~\ref{fig:sim1d_kymo}(B). In this case, the pattern propagates at the same speed as the wave speed $\alpha$, since pattern formation is restricted to behind the moving front. %, so it cannot propagate at the faster natural rate. 
Here new spikes form via insertion between the moving front and the previous leading spike, and remain mostly static after their formation. In contrast, with an even slower wave speed, $ \alpha \ll \nu$, as in Fig.~\ref{fig:sim1d_kymo}(C), there is a leading spike that travels at the same speed, $\alpha$, as the moving front, and new spikes form behind it by splitting off the leading spike. After the split, the leading spike continues to move in tandem with the moving boundary.
For an intermediate range of wave speeds $0.2 < \alpha < 0.3$, we observe a mixture of behaviours, where both peak insertion and splitting occurs. We therefore define $\alpha_c \approx 0.25$ to be a threshold that approximately separates the two cases.

%(A) $\alpha=2$ ; (B) $\alpha=0.75$ ; (C) $\alpha=0.1$ ; (D) fixed domain

We denote these observed behaviours cases A, B, and C, and summarize them in Table~\ref{tab:cases}, including the fixed domain case (case D) for comparison. %These names are the same as the labels of their corresponding diagrams in Fig.~\ref{fig:sim1d_kymo}, and subsequent figures.
In all of these cases, the final pattern is the same, %\mycomment{do we need the plot for this? I'm inclined toward no} 
and we observe sequential formation of spikes from left to right \revision{(except near the right-hand boundary, where potential boundary effects may disturb the pattern)}, either at speed $\nu$ or $\alpha$, whichever is slower. However, there are important differences in the transient behaviours.
In Fig.~\ref{fig:sim1d_kymo}(A), we have a part of the domain that momentarily remains at the homogeneous steady state after it has become capable of supporting a pattern, whereas in Fig.~\ref{fig:sim1d_kymo}(B,C) the pattern always fills the entirety of $\Omega'$. 

The results for the CDIMA model are largely similar to the observations described above for the Schnakenberg model. We were able to estimate $\nu \approx 1.5564$ for the CDIMA model for the parameter values in Eq.~\eqref{eqn:cdima_params}, which separates case A from cases B and C in a similar way.
One key difference is that in both cases B and C, the CDIMA model exhibits peak splitting behaviour similar to  case C of the Schnakenberg model. It does not appear to exhibit peak insertion behaviours for any of the tested value of $\alpha$ within the relevant range.
This means that we can only observe one, instead of two, distinct behaviours for the range  $0 < \alpha < \nu$, in the CDIMA model. 

\newpage
\begin{table}[H]
    \centering
    \begin{tabular}{c||c|c|c}
        Cases & Range of $\alpha$ & Behaviour in 1D simulations & Behaviour in noisy 2D simulations \\
        \hline
        \hline
        A & $\alpha > \nu$ & \specialcell{Peaks propagate at speed $\nu$ \\(noiseless, Fig.~\ref{fig:sim1d_kymo}(a))\\ Disordered peak formation \\(noisy, Fig.~\ref{fig:sim1d_kymo_noisy}(a)) } & \specialcell{Labyrinthine patterns \\ (Fig.~\ref{fig:sim2d_noisyinit},\ \ref{fig:sim2d_wavyinit}, first column)}\\
        \hline
        B & $0 < \alpha_c \ll \alpha < \nu$ & \specialcell{Peak insertion\\ (Fig.~\ref{fig:sim1d_kymo}(b),\ \ref{fig:sim1d_kymo_noisy}(b))}& \specialcell{Stripes parallel to moving front\\ (Fig.~\ref{fig:sim2d_noisyinit},\ \ref{fig:sim2d_wavyinit}, second column)}\\
        \hline
        C & $0 < \alpha \ll \alpha_c$ & \specialcell{Peak splitting\\ (Fig.~\ref{fig:sim1d_kymo}(c),\ \ref{fig:sim1d_kymo_noisy}(c))}& \specialcell{Stripes transverse to moving front\\ (Fig.~\ref{fig:sim2d_noisyinit},\ \ref{fig:sim2d_wavyinit}, fourth column)}\\
        \hline
        D & \specialcell{$\alpha=0$ \\(fixed domain)}& \specialcell{Peaks propagate at speed $\nu$ \\(noiseless, Fig.~\ref{fig:sim1d_kymo}(d))\\ Disordered peak formation \\(noisy, Fig.~\ref{fig:sim1d_kymo_noisy}(d))}  & \specialcell{Labyrinthine patterns\\ (Fig.~\ref{fig:sim2d_noisyinit},\ \ref{fig:sim2d_wavyinit}, fifth column)}\\
    \end{tabular}
    \caption{Summary of the four cases identified for the Schnakenberg model in Section~\ref{sec:sim1d}, and their behaviours observed in simulations on one-dimensional domains (without on-going noise (Fig.~\ref{fig:sim1d_kymo}), and with on-going noise (Fig.~\ref{fig:sim1d_kymo_noisy})), and two-dimensional domains, with on-going noise (Fig.~\ref{fig:sim2d_noisyinit}, \ref{fig:sim2d_wavyinit}).
    Two critical thresholds separate the cases: $\nu$, the natural speed of pattern propagation defined in Section~\ref{sec:sim1d}, and $\alpha_c$, which was estimated using a parameter sweep.}
    \label{tab:cases}
\end{table}

Next, we consider the impact of on-going noise, as discussed in Section~\ref{sec:models}.
In case A, as well as the fixed domain case D, we would expect this noise to trigger the formation of new spikes in the region of $\Omega'$ where the system would remain close to the homogeneous steady state in the absence of noise, disrupting the orderly sequential formation of spikes. On the other hand, in cases B and C, we do not expect this additional noise to have a major effect. This is because the pattern always fills the entirety of $\Omega'$, 
so the noise decays in $\Omega'$ because the multi-spike solution is stable, and it also decays in $\Omega \setminus \Omega'$ due to the spatially uniform steady state being stable in that subdomain.

\begin{figure}[H]
    \centering
    \includegraphics[width=1\textwidth]{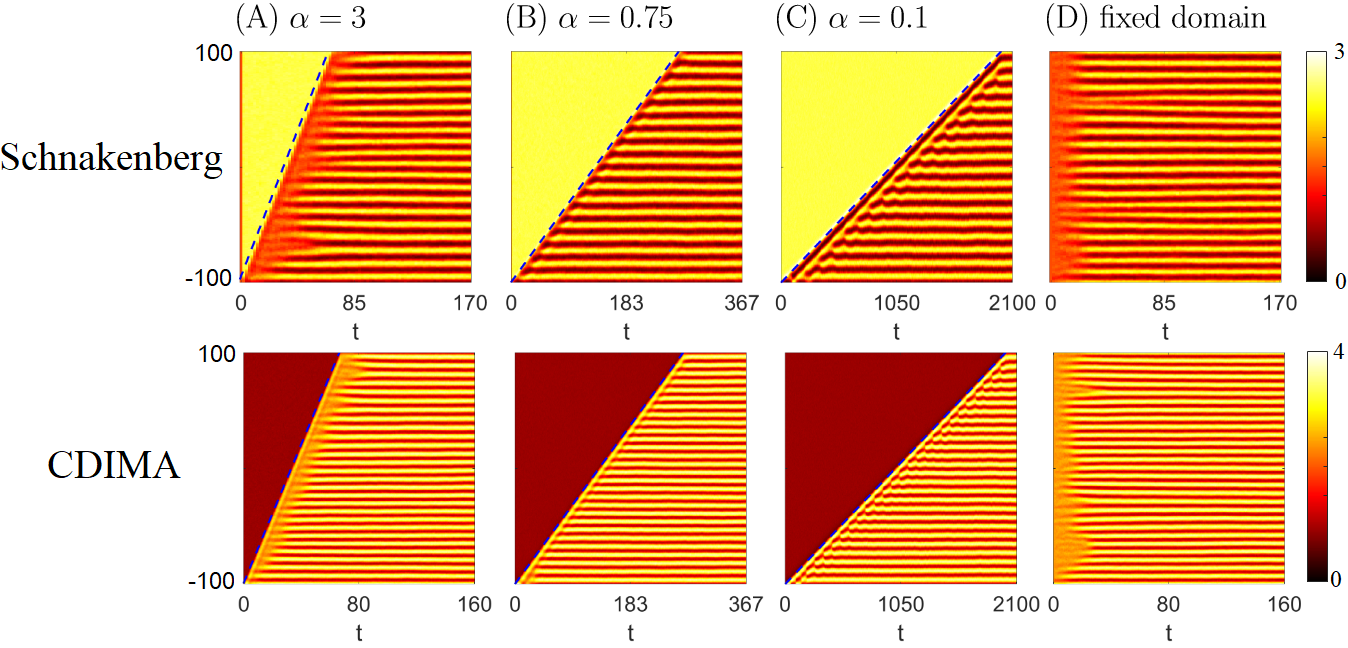}
    \caption{Solution of the Schnakenberg and CDIMA models, with the same parameters, wave speeds and initial conditions as in Fig.~\ref{fig:sim1d_kymo}, except with on-going noise. The blue dashed lines indicate the boundary of $\Omega'(t)$.
    %Corresponding animations are provided at \url{imgur.com/a/0dy3kPY}. 
    Observe that in (A) and (D), the spikes no longer form in an orderly sequence, whereas (B,C) are virtually identical to the corresponding cases in Fig.~\ref{fig:sim1d_kymo}.}
    \label{fig:sim1d_kymo_noisy}
\end{figure}

The numerical results from our simulations confirm our expectations.
In Fig.~\ref{fig:sim1d_kymo_noisy}, we observe that, indeed, the sequence of spike formation is disrupted in case A. Once again, the behaviour in this case closely resembles the behaviour on a fixed domain (D), where the pattern fills the patterning subdomain $\Omega'$, which coincides with entire domain $\Omega$ in the latter case. 
In contrast, the behaviours in cases B and C remain virtually unchanged compared to the noiseless case in Fig.~\ref{fig:sim1d_kymo}, as predicted.
\revision{In the presence of on-going noise, spike formation is initiated everywhere in the patterning-competent subdomain $\Omega'$ simultaneously, instead of spreading outward from a single point at speed $\nu$. Nonetheless, $\nu$ remains the separator of the distinct behaviours in cases A and B, since the size of $\alpha$ relative to $\nu$ determines whether the pattern always fills $\Omega'$.}

The same behaviours are observed for the CDIMA model in the noisy case, except for the difference noted before, where peak splitting is observed instead of peak insertion in case B. 

From the numerical simulations in one dimension, we are able to draw the following conclusions for the Schnakenberg model. First, if the wave speed, $\alpha$, which determines the rate of expansion of the patterning subdomain, is faster than the natural speed of pattern propagation, $\nu$ (case A), then there will be little interaction between the moving front and the pattern forming mechanism, in which case we expect similar behaviour 
to the case where the whole domain can undergo DDI. 
However, for $\alpha< \nu$ we can have very different behaviours. A high wave speed, $\alpha$, leads to the formation of new spikes via the insertion of peaks behind the moving front (case B), whereas for a sufficiently low wave speed, $\alpha$, the spikes form via the splitting of a leading, moving spike (case C). Moving forward into two dimensions, these observations suggest that in the presence of on-going noise, case A will lead to disorganized patterns, matching the behaviour on a fixed domain, whereas cases B and C may lead to the formation of more organized, but possibly distinct, patterns, following the wave of competency. These predictions apply to the CDIMA model as well, except we might expect fewer differences in behaviour between cases B and C.

\subsection{Simulations in two spatial dimensions}\label{sec:sim2d}

With the knowledge gained from the four cases observed in one dimension, we now investigate whether we observe the same distinct behaviours in two dimensions, and if the cases in one and two dimensions have any correspondence between them.
We run all simulations long enough for $\Omega'$ to expand to cover the entire domain $\Omega$, plus a further 200 time units. By numerical experimentation, we found that this is sufficient time for the system to reach a steady state.
For comparison, we also run simulations where we set $\Omega'=\Omega$ (that is, the whole domain is capable of undergoing DDI immediately), in which case we run the simulations until $t=200$.
We present only the final profile for $u$ in the two-dimensional simulations, since the profile for $v$ is either in- or out-of-phase with respect to $u$, depending on the kinetic terms, so the alignment and orientation of the stripes will be the same for both morphogens.
We will first briefly mention the observations in the noiseless case, then investigate how the results change with the addition of on-going noise.

For both the Schnakenberg and CDIMA models, we use the domain described in Section~\ref{sec:models}, 
with the parameters from Eqs.~\eqref{eqn:schnackenberg_params} and \eqref{eqn:cdima_params}, and we initialize the system by 
\revision{perturbing $u$ on the left-hand boundary by setting $u(x=-L,y)=2u_*$.
This choice is consistent with the initial condition used in the one-dimensional simulations, which can be recovered by taking any horizontal cross-section with $y$ constant.}
\revision{By observing the simulation on a fixed domain (Fig.~\ref{fig:sim2d_nonoise}, see supplementary materials for the corresponding animation), we found the same value of $\nu$ as we did in one dimension.}
Without on-going noise, we found that for the range of wave speeds $0.1 \leq \alpha \leq 2$, the system always displays vertical stripes parallel to the moving boundary for both models (Fig.~\ref{fig:sim2d_nonoise}). 
\revision{Even if we consider a larger perturbation to the initial condition, e.g. adding uniformly distributed noise in $(-0.3,0.3)$ to each point in the domain, the behaviour remains the same.}
The reason for this result is that since initially $\Omega'$, the subdomain capable of supporting Turing patterns, is very small, the noise in the initial condition decays, similar to what we have found in Section~\ref{sec:sim1d}, leaving the system essentially at its homogeneous steady state in $\Omega \setminus \Omega'$. This, combined with the regular square mesh employed in the numerical simulations, means that there is no symmetry breaking mechanism in the $y-$direction. Consequently, the patterns we obtain are simply the multi-spike pattern in the $x-$direction, but constant in the $y-$direction. This reduces the system back to the one-dimensional case. While we do observe differences in the transient behaviour as $\alpha$ changes, analogous to the observations in Section~\ref{sec:sim1d}, the final pattern is always the same.

\begin{figure}[H]
    \centering
    \includegraphics[width=0.5\textwidth]{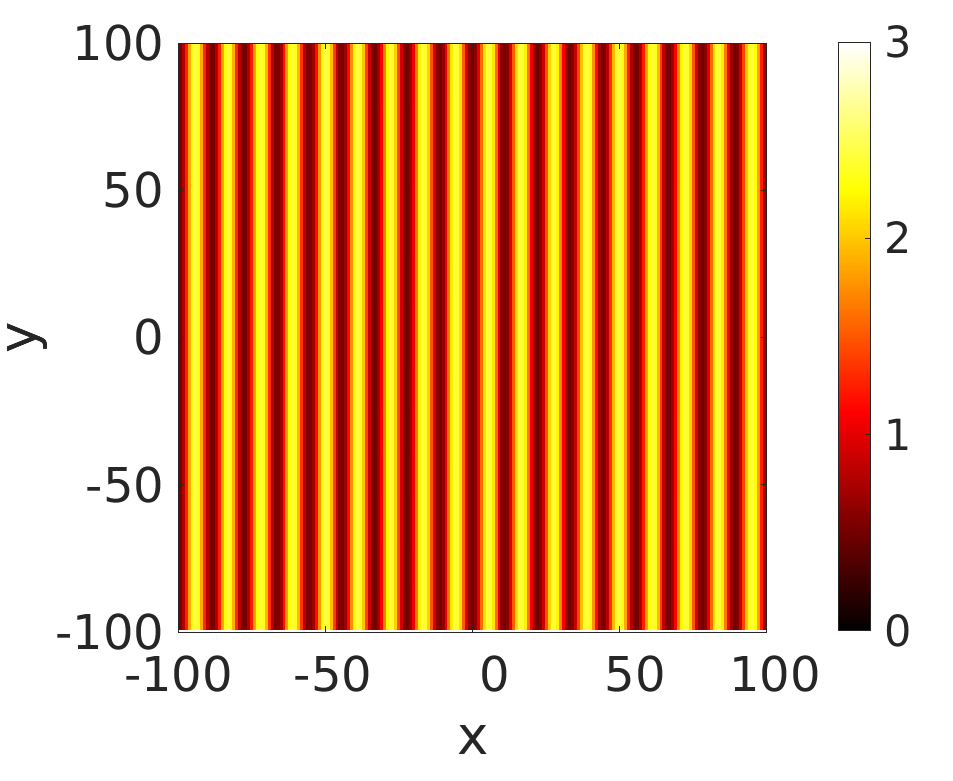}
    \caption{The final pattern produced with the Schnakenberg model on a two-dimensional domain with parameters from Eq.~\eqref{eqn:schnackenberg_params} and $b=1.4, \  \alpha=0.75$, initialized by perturbing the homogeneous steady state at each mesh point, without any on-going noise. The resulting pattern is a series of perfectly vertical stripes. Similar results were obtained for other values of $\alpha$ in the range $[0.1,2]$, \revision{as well as on a fixed domain,} and for the CDIMA model.}
    \label{fig:sim2d_nonoise}
\end{figure}

We also attempted to recover the results of \citep{konow2019TuringPatternsRadially} for the CDIMA model, which uses a circular domain that grows radially. The simulations without on-going noise were only able to produce circular rings parallel to the growing boundary, and not the other patterns observed in \citep{konow2019TuringPatternsRadially}. Once again, this is an artifact of the regularity of the mesh used; \citep{konow2019TuringPatternsRadially} used a finite element method with an irregular mesh, which naturally acts as a symmetry-breaking mechanism.
However, simulating the model with on-going noise, we can reproduce all patterns from \citep{konow2019TuringPatternsRadially} (see \ref{apx:circular_domain}).

\begin{figure}[H]
    \centering
    \includegraphics[width=\textwidth]{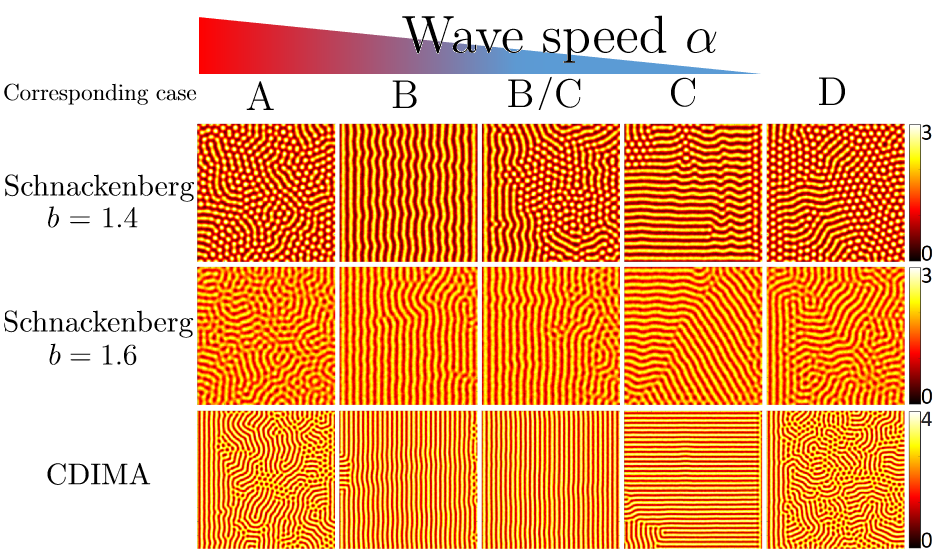}
    \caption{Examples of the steady state patterns produced by the Schnakenberg model~\eqref{sys:schnackenberg}, and the CDIMA model~\eqref{sys:cdima}, with choice of domain as given in Eq.~\eqref{eqn:domain}, and the wave speed decreasing from left to right. The last column shows the fixed domain case. 
    For both models, $\alpha = 3, 0.75, 0.5, 0.2$.
    %Only the profile of $u$ is shown, the profile for $v$ is in-phase to $u$ for CDIMA model and out-of-phase for Schnakenberg.
    Notice that in the first column, corresponding to case A in Table~\ref{tab:cases}, the pattern begins as vertically aligned stripes on the left side, but quickly transitions to a labyrinthine pattern as pattern formation lags behind the wave of competency.
    The second column corresponds to case B, and the patterns mostly take the form of vertical stripes. 
    The third column, corresponding to an intermediate regime between cases B and C, exhibits a mixture of the behaviours from the second and fourth columns, while the fourth column (corresponding to case C) contains regions where the stripes are slanted or horizontal.
    Finally, the last column shows the labyrinthine patterns produced for the case where $\Omega'=\Omega$. 
    %The animations showing the evolution of these patterns are provided at \url{imgur.com/a/u2b7qaV}.
    }
    \label{fig:sim2d_noisyinit}
\end{figure}

For the simulations with on-going noise, we first simulate the system of PDEs~\eqref{sys:generic_rd} with both sets of kinetics, initialized at the homogeneous steady state. We add the same perturbation \revision{on the left-hand boundary to the initial condition as before for consistency, however we note that the specific form of perturbation makes little difference in this case, and we recover the same final patterns if the initial condition is perturbed in a different way, e.g. by adding a uniformly distributed noise at each mesh point in the domain. As explained before, we expect the initial condition to have little influence on the final pattern since it almost entirely decays away before the onset of pattern formation.}
The results are shown in Fig.~\ref{fig:sim2d_noisyinit}, where the rows correspond to the different kinetics, and the columns are arranged in the order of decreasing wave speed, $\alpha$.
In the first column, which is the case $\alpha > \nu$, corresponding to case A in Table~\ref{tab:cases}, we can observe that the pattern initially takes the form of vertical stripes, but quickly transitions to become disordered and labyrinthine, matching the behaviour on a fixed domain (case D), as shown in the fifth column. This is reminiscent of the one-dimensional behaviour for the same case in Fig.~\ref{fig:sim1d_kymo_noisy}(A), where the sequential formation of spikes is disrupted and the pattern becomes disordered, matching the behaviour in Fig.~\ref{fig:sim1d_kymo_noisy}(D).

In the second column of Fig.~\ref{fig:sim2d_noisyinit}, we have \revision{$0 < \alpha \lesssim \nu$}, corresponding to case B in Table~\ref{tab:cases}. With a moderately high $\alpha$, for both the CDIMA and Schnakenberg models, multiple repeated simulations with different random seeds %(for both the noise for the initial conditions and the on-going noise) 
produced only stripes aligned vertically, parallel to the moving boundary of the patterning subdomain representing the wave of competency.
This strongly suggests that vertical stripes are the only pattern possible in this case.

For sufficiently small values of $\alpha$, as in the fourth column, corresponding to case C in Table~\ref{tab:cases}, the stripes may be aligned horizontally, or at a slanted angle (most clearly visible in the Schnakenberg $b=1.4$ pattern). The CDIMA model tends to form a few, large regions where the stripes are in alignment, where there are two main regions, one with vertical stripes, and the other horizontal stripes. In comparison, the Schnakenberg model tends to exhibit a greater number of such regions. For both models, the final pattern is highly dependent on the noise. The number and location of these regions containing aligned stripes, as well as the direction of stripes within them, can be completely different when a different random seed is used.
The third column appears to exhibit a mixture of its two neighbouring columns. The Schnakenberg model produced some slanted stripes, but not as many as in the fourth column.

We also observe in Fig.~\ref{fig:sim2d_noisyinit} that the Schnakenberg model with $b=1.4$ is much more prone to producing spots interspersed between stripes than the same model with $b=1.6$, or the CDIMA model. However, with an expanding patterning subdomain, the model tends to produce more aligned stripes compared to a fixed domain.

These results suggest that the vertical stripe pattern is the only stable pattern in case B (moderately high $\alpha$), while in case C (low $\alpha$), we observe multi-stability where stripes aligned at a variety of angles are possible, including horizontal stripes. 
This leads us to ask whether the horizontal stripes are stable at low $\alpha$, and if so,  how high does $\alpha$ need to be for the horizontal stripes to become unstable?

To answer this question, we simulate the same models, but using initial conditions chosen as horizontal stripes in the form of cosines, with the amplitude and period set to the estimated values from the final steady state of the previous simulations. 
Specifically, we set 
\begin{equation}
u(x,y,0)=m_u+A_u\Cos{qy}, \quad v(x,y,0)=m_v+A_v\Cos{qy}, \label{eqn:horizontal_stripes}
\end{equation}
where $m_u,A_u,m_v,A_v,$ and $q$ are estimated to match the patterns in the second column of Fig.~\ref{fig:sim2d_noisyinit} by measuring the average wave length and amplitude of the final patterns.
For the CDIMA model, we used \[m_u=2.41,\quad A_u=1.16,\quad m_v=6.67,\quad A_v=0.71,\quad q=0.91.\] 
For the Schnakenberg model, we used \[m_u=1.45,\quad A_u=0.98,\quad m_v=0.65,\quad A_v=-0.20,\quad q=0.57.\]
\revision{This set of initial conditions, unsurprisingly, leads to the same pattern as before, for the aforementioned reason that, with our choice of $\rho(t)$, the initial condition decays to the homogeneous steady state before pattern formation can take place. Therefore, instead of the form of $\rho(t)$ as defined in Eq.~\eqref{eqn:domain}, we take
\[\rho(t)=\min(-L_x+L_0+\alpha \max(t-t_0,0), L_x), \quad L_0=0.2L_x, \quad t_0=50,\]
meaning that the wave of competency remains at $L_0$ up to time $t_0$, allowing a pattern to develop inside $\Omega'$ before the wave begins to advance at constant speed $\alpha$.
}

\newpage
\begin{figure}[H]
    \centering
    \includegraphics[width=\textwidth]{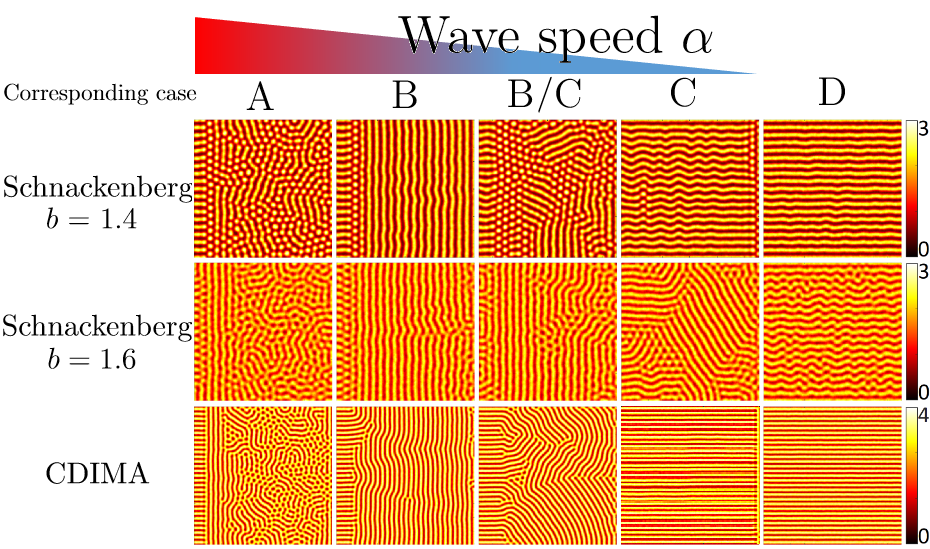}
    \caption{Examples of the steady state patterns produced by the Schnakenberg model~\eqref{sys:schnackenberg}, and the CDIMA model~\eqref{sys:cdima}, where the wave speed decreases from left to right. The parameter values and wave speeds are the same as in Fig.~\ref{fig:sim2d_noisyinit}, except the initial conditions are set to horizontal stripes, as described in the text. 
    In the first column, for high $\alpha > \nu$, we observe disordered labyrinthine patterns, as before.
    In the second column, at moderately high $\alpha$, the pattern abruptly switches to vertical stripes for both models, once again suggesting that this is the only stable pattern at this wave speed. 
    In the third column, a medium value of $\alpha$ leads to a mixture of behaviours. In the CDIMA model, the stripes curve smoothly to become more vertical, while the Schnakenberg model with $b=1.4$ forms an abundance of spots, which separates regions of aligned stripes. However, for $b=1.6$, a similar pattern without spots is produced by the Schnakenberg model.
    In the fourth column, where $\alpha$ is low, we see that horizontal stripes persist for both models, although in the Schnakenberg model the stripes may bend away at a slanted angle.
    Finally, in the fifth column, unsurprisingly, we observe that the horizontal stripes persist in the fixed domain case. 
    %The corresponding animations can be found at \url{imgur.com/a/SiasbMa}.
    }
    \label{fig:sim2d_wavyinit}
\end{figure}

The results are shown in Fig.~\ref{fig:sim2d_wavyinit}. 
For $\alpha > \nu$ (first column, corresponding to case A), we observe the same disordered labyrinthine patterns as we did in Fig.~\ref{fig:sim2d_noisyinit}, since the initial stripe pattern in $\Omega \setminus \Omega'$ quickly decays to the homogeneous steady state, essentially recreating the same initial conditions as in the earlier simulations.
For \revision{$0 < \alpha \lesssim \nu$} (second column, case B), both models produce vertical stripes, and the transition from horizontal stripes to vertical stripes is abrupt. Some spots appear near the transition, but in both  the CDIMA model and the Schnakenberg model with $b=1.6$, the spots are eventually absorbed into new stripes.

For  $\alpha \ll \nu$ (fourth column, case C), the horizontal pattern persists for the CDIMA model and for the Schnakenberg model with $b=1.4$. For the Schnakenberg model with $b=1.6$, the stripes bend away in some parts of the domain and align at an angle, but nonetheless horizontal stripes dominate. Whether or not this bending occurs is noise-dependent and differs between realisations. %\mycomment{I mean there is a chance that this happens if the noise hits just right, not about the level of the noise. How to say this?}
For medium $\alpha$ (third column), in both the CDIMA model and the Schnakenberg model with $b=1.6$, the stripes curved away smoothly from horizontal to become more vertical. For the Schnakenberg model with $b=1.4$, the curved stripes are replaced with spots.

For the fixed domain case (fifth column, case D), the on-going noise (recall that the noise has magnitude $\mu=0.01$) was unable to destabilize the initial stripe pattern. This is the only scenario where we obtain different outcomes in cases A and D. 
However, for the fixed domain case, the pattern can be destroyed if the noise strength is increased sufficiently, see \ref{apx:high_noise} for details.
Note that in both Fig.~\ref{fig:sim2d_noisyinit} and Fig.~\ref{fig:sim2d_wavyinit}, some of the stripes are not perfectly straight, but rather are a little curved. This is a consequence of the noise added at each time step.

\begin{figure}
    \centering
    \includegraphics[width=0.5\textwidth]{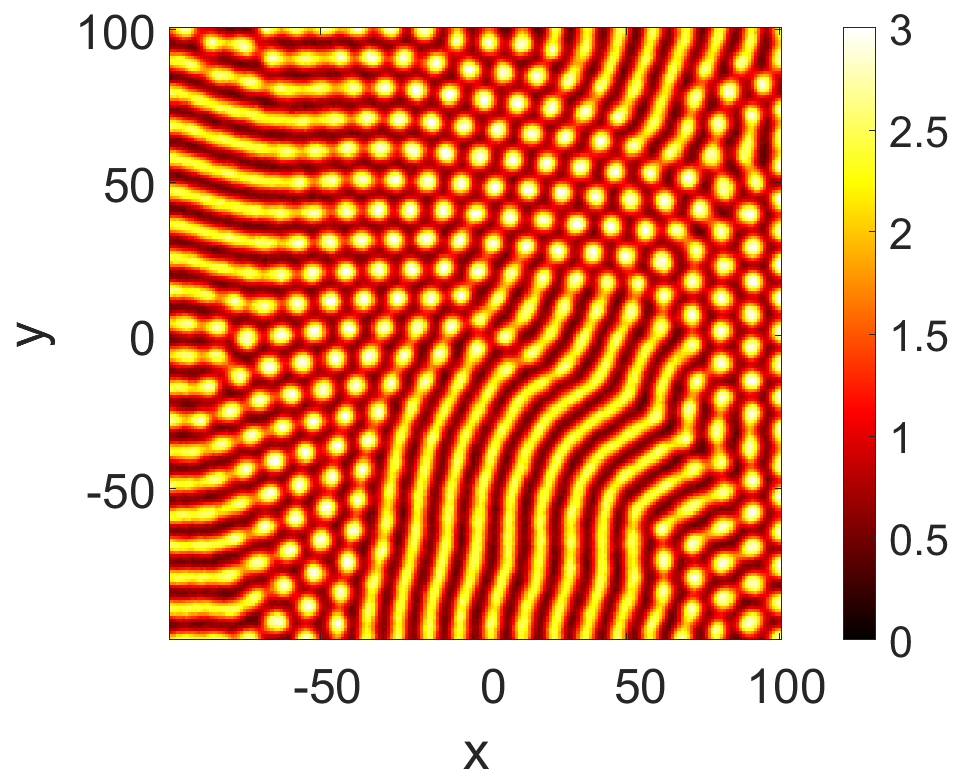}
    \caption{
    \revision{Final pattern for the Schnakenberg model with $b=1.4$ and $\alpha=0.05$, with initial conditions set to horizontal stripes, as in Fig.~\ref{fig:sim2d_wavyinit}. Note that the pattern is mostly disordered.}
    }
    \label{fig:sim2d_verylow_alpha}
\end{figure}

\revision{
We also note that with an extremely low speed, $\alpha$, below the value used in case C, the timescale for pattern formation is much shorter than the timescale for the expansion of the competent subdomain $\Omega'$, so the dynamics inside $\Omega'$ can be seen as quasi-static. Therefore, the effect of the wave of competency decouples from the pattern formation process. Over the long time it takes for the wave of competency to traverse the full domain, the pattern will eventually be destabilised by the on-going noise and become disordered.
This effect is more easily seen for the Schnakenberg model with $b=1.4$ since, compared to the CDIMA model or the Schnakenberg model with $b=1.6$, it has a greater tendency to exhibit stripes that break up into spots, and therefore the stripe pattern is more easily destabilised by the on-going noise. An example is given in Fig.~\ref{fig:sim2d_verylow_alpha} where, for a very low value of $\alpha$, the initial horizontal stripes do not persist very far beyond the left-hand boundary, and the final pattern consists of a mixture of disordered spots and stripes, similar to the fixed domain case with uniform initial conditions (Fig.~\ref{fig:sim2d_noisyinit}, final column).
Since we are mostly interested in cases where the two timescales are comparable, we will not consider this case further.
}

These simulations show a clear correspondence with the cases in one dimension. 
First, for case A \revision{($\alpha < \nu$; first column in Figs.~\ref{fig:sim2d_noisyinit},\ref{fig:sim2d_wavyinit}), for both sets of initial conditions,} we obtain labyrinthine patterns similar to case D \revision{with homogeneous initial conditions (Fig.~\ref{fig:sim2d_noisyinit}, final column)}. This is analogous to the results observed in one dimension, where similar transient behaviours were observed in cases A and D.
For case B  \revision{$(0 < \alpha \lesssim \nu)$}, our simulations suggest that the only stable pattern is vertical stripes parallel to the moving front, whereas for case C $(\alpha \ll \nu)$, patterns that are transverse to the moving front, including horizontal and slanted stripes, appear to be stable.

Moreover, for the Schnakenberg model, we found that the switch between cases B and C happens at approximately the same critical value of $\alpha$ in both one and two dimensions, suggesting that the distinction between peak insertion and peak splitting in one dimension may correspond to the distinction between vertical stripes and transversal stripes in two dimensions.
For the CDIMA model, despite our observations that it exhibits the same qualitative behaviour in cases B and C in one dimension, we find it exhibits different outcomes in two dimensions that largely agree with the behaviour of the Schnakenberg model, where the preferred directionality of the stripes changes.
While this is a surprising contrast to the one-dimensional behaviours, it is consistent with the results in \citep{konow2019TuringPatternsRadially}, where distinct patterns analogous to those we have observed at different values of $\alpha$ were also observed. 

While the critical threshold between cases A and B appears to be at $\alpha=\nu$, the threshold between cases B and C is much less clear. We will explore how these two cases can be distinguished, at least for the Schnakenberg model, in the next section.

\subsection{Single stripes on a narrow two-dimensional domain}\label{sec:sim2d_narrow}

The simulations so far on one- and two-dimensional domains have allowed us to identify three distinct patterning regimes under the effect of a wave of competency, denoted by cases A, B, and C. While a promising candidate for the critical value of $\alpha$ separating cases A and B has been found, we still need to  delineate cases B and C. 

\begin{figure}
    \centering
    \includegraphics[width=0.4\textwidth]{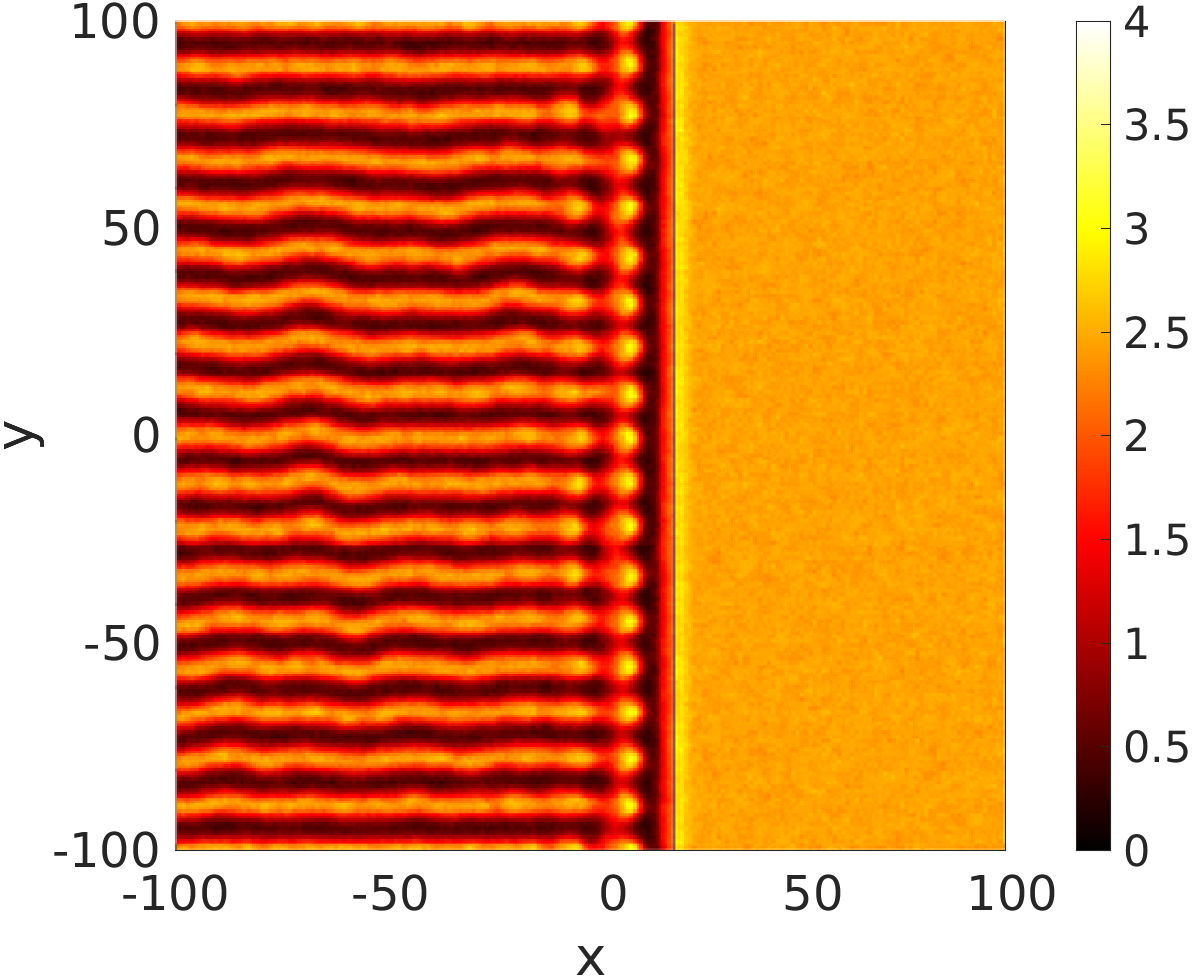}
    \caption{
    \revision{Snapshot of the simulation corresponding to the pattern presented in Fig.~\ref{fig:sim2d_wavyinit}, for the Schnakenberg model with $b=1.4, \alpha=0.2$, taken at $t=550$. Here the spot structure at the head of each stripe is clearly visible. }
    }
    \label{fig:sim2d_snapshot}
\end{figure}

By examining animations of the simulations in two dimensions, we  observe that during the formation of stripes transversal to the moving front (including horizontal and slanted stripes), there is usually a spot-like structure trailing right behind the front, with the rest of the stripe extending behind it. \revision{This structure can be seen in the snapshot presented in Fig.~\ref{fig:sim2d_snapshot}.} Since this behaviour is observed for horizontal stripes and slanted stripes, we will classify both of these orientations  together as the same type of pattern henceforth. 
In contrast, a vertical stripe forms spontaneously when the space between existing stripes and the front becomes sufficiently large. This behaviour is common to both models.
It turns out that, for the Schnakenberg model, the key to distinguishing cases B and C is to look closely at the structure of the stripe as it lengthens.

\begin{figure}[H]
    \centering
    \includegraphics[width=\textwidth]{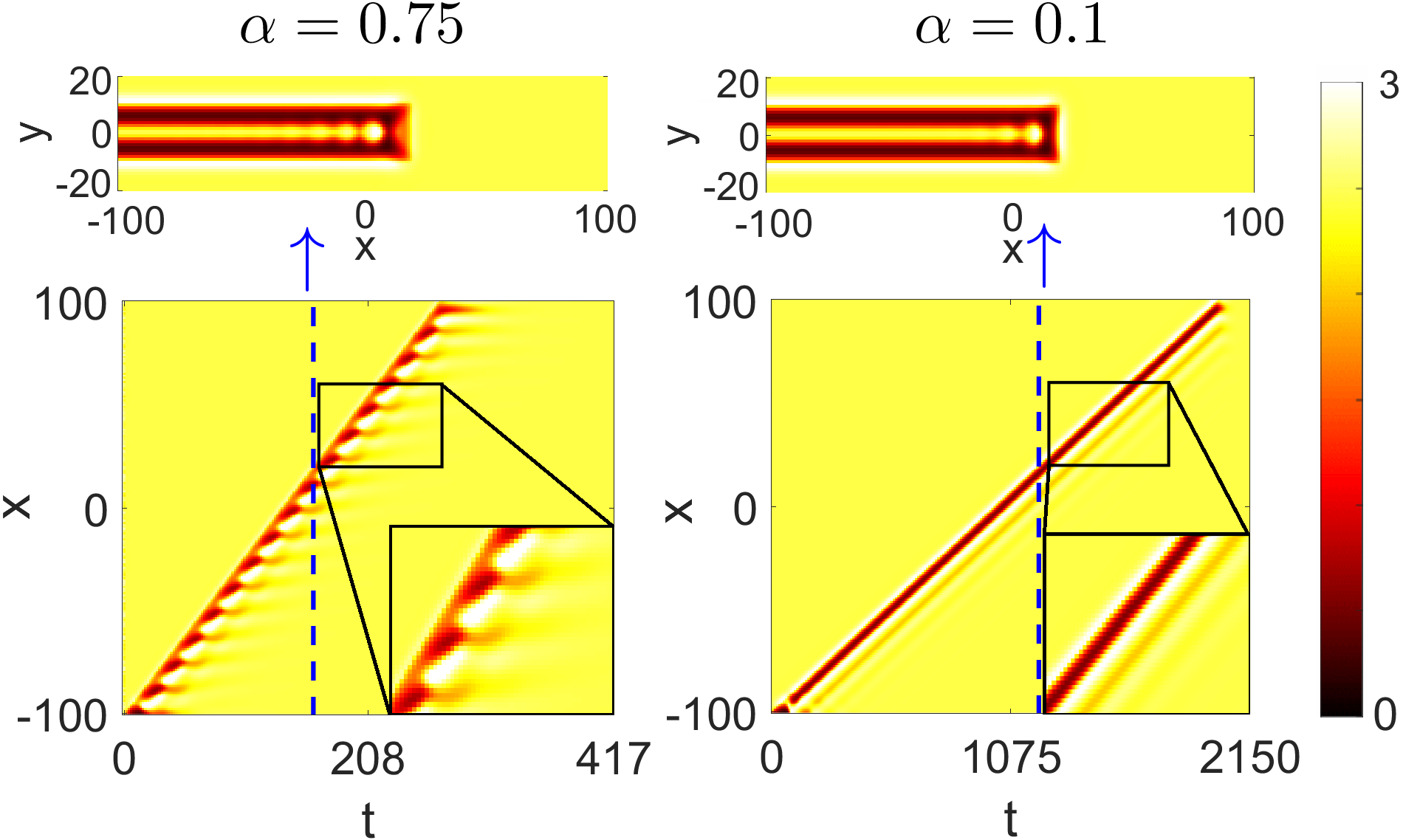}
    \caption{
    Simulations of the Schnakenberg model (Eq.~\eqref{sys:schnackenberg}) with $b=1.4$ on a narrow patterning subdomain as in Eq.~\eqref{eqn:narrow_domain}, represented with kymographs of $u$ along the cross-section $y=0$, and a snapshot of the two-dimensional stripe during its formation. The vertical blue line indicates where the snapshot was taken.
    For $\alpha=0.75$, which corresponds to case B, we can observe the formation of a series of spots via insertion between the moving boundary and the trailing stripe. They persist for only a short time before being absorbed into the stripe.
    In contrast, for $\alpha=0.1$, which corresponds to case C, we observe a moving spot travelling at the same speed as the moving boundary, forming a travelling wave solution along the cross-section $y=0$.
    %The corresponding animations are provided at \url{imgur.com/a/haAoUZg}.
    }
    \label{fig:sim2d_narrow}
\end{figure}

To do so, we modify the subdomain capable of undergoing a DDI to be a narrow rectangle that elongates over time. That is, we now consider
\begin{equation}
\Omega' = [-L_x,\rho(t)] \times [-0.1L_y,0.1L_y] , \label{eqn:narrow_domain}
\end{equation}
where $\rho(t)$ is as before.
We will focus on the Schnakenberg model with $b=1.4$, then describe how the behaviour of the CDIMA model differs. 
The narrow width of the patterning subdomain means that no more than one horizontal stripe can be supported. We also found that the stripe does not break up into a row of spots for either model, and the final pattern is a single horizontal stripe for all $\alpha$ in the range $[0.1,0.75]$. Typical snapshots of the pattern can be found in Fig.~\ref{fig:sim2d_narrow}.
%Since we cannot distinguish cases B and C from the final pattern, we must focus on the transient behaviour during the process of stripe formation.
We will focus on the transient behaviour during the process of stripe formation, just like we have done in one dimension.

In this case, on-going noise is not necessary to capture the interesting part of the behaviours. In fact, we obtain the same qualitative results regardless of the presence of the on-going noise. This is not surprising since, just as discussed for the one-dimensional case (Section~\ref{sec:sim1d}), with $\alpha < \nu$ the pattern always fills $\Omega'$ in the noiseless simulations. Without any freedom to extend the pattern in the $y$--direction, the noise has little effect on the pattern.
Therefore, we will only present the results from simulations without noise.

We illustrate the transient behaviour with a kymograph of $u$ along the horizontal cross-section at $y=0$, along the ridge of the single stripe. The results are presented in Fig.~\ref{fig:sim2d_narrow}. We have also included a snapshot of the stripe during its formation for visualization.

For sufficiently slow wave speeds, $\alpha \ll \alpha_c$, for some critical threshold $\alpha_c$, there is a single leading spot right next to the moving boundary, as in the earlier simulations in two dimensions, which moves in tandem with the front, with the rest of the stripes trailing behind it, remaining a constant distance away from the leading spot. 
This can be seen in Fig.~\ref{fig:sim2d_narrow} (right), where the two white diagonal lines, representing the location of the front and the leading spot, are parallel to each other.
The solution along the horizontal cross-section at the middle of the domain, $y=0$, is effectively a travelling wave, moving at the same speed, $\alpha$, as the moving boundary.

In contrast, for faster wave speeds, $\alpha \gg \alpha_c$, the leading spot remains stationary, and the trailing stripe expands by engulfing the spot. New leading spots are inserted periodically between the moving boundary and the previous leading spot, as the space between them becomes sufficiently large. We can see this in Fig.~\ref{fig:sim2d_narrow} (left), where the locations of all local maxima stay roughly fixed. 
We performed a parameter sweep, simulating the system for a range of values of $\alpha$, and found that travelling-wave-like behaviour is observed for $\alpha < 0.31$, while spot-insertion behaviour occurs for $\alpha > 0.45$. 
For the intermediate values of $\alpha$ lying between these values, we observe a mixture of behaviours.
Our analysis allows us to make a crude estimate of the threshold $\alpha_c \approx 0.37$  for the Schnakenberg model in this context. Notice that this value for $\alpha_c$ is higher than what we found with simulations on one-dimensional domains. This difference is likely due to the diffusion in the $y$--direction having a non-trivial impact on the dynamics at the cross-section.
This value of $\alpha_c$  serves as a rough guide to delineate cases B and C for the Schnakenberg model. In a sufficiently wide two-dimensional domain (as in Section~\ref{sec:sim2d}), we expect stripes parallel to the moving front to form if $\alpha \gg \alpha_c$, and stripes transverse to the moving front to arise if $\alpha \ll \alpha_c$.
Just as we have seen in Section~\ref{sec:sim2d}, the boundary between the two cases is not clear-cut, as we observe mixed behaviours for $\alpha$ near $\alpha_c$ in the third column.

For the CDIMA model, in the narrow two-dimensional domain described in Eq.~\eqref{eqn:narrow_domain}, we observe the same travelling wave-like behaviour in both cases B and C, which is consistent with the observations made in one dimension.

\section{Discussion}\label{sec:discuss}

The aim of this work was to explore the impact of a wave of competency upon DDI models of pattern formation,
taking as exemplar kinetics the Schnakenberg and CDIMA models. 
Through extensive numerical exploration, we have shown that the speed of the wave of competency can be used to select the resulting pattern. Specifically, the alignment and directionality of striped patterns in two dimensions can be selected to be either parallel or perpendicular to the wave of competency by appropriate choice of the wave speed. Moreover, this phenomenon holds true for both the CDIMA and Schnakenberg kinetics, with minor differences in the model behaviours in one dimension. 
The agreement between the two models suggests that our results for two-dimensional behaviours may possibly be generalized to a larger class of models capable of exhibiting a DDI, i.e. that the phenomenon does not depend on the specific form of the reaction terms, nor on whether the morphogens are in- or out-of-phase. 
Our observations are robust to the addition of noise, which gives us confidence that the results on stripe alignment and directionality are robust. 

We have observed three main patterning regimes on expanding patterning subdomains with respect to the value of the wave speed $\alpha$. These cases (A,B,C), along with the behaviour on the fixed domain (case D) for comparison, are summarized in Table~\ref{tab:cases}. We have defined the natural speed of pattern propagation, $\nu$, as the natural speed at which a pattern spreads from a spatially localised perturbation on a fixed domain. Our first main conclusion, which holds for both models, is that in order for stripe patterns in two dimensions to align in the presence of noise, the wave speed must be slower than the speed at which the pattern propagates naturally, that is $\alpha<\nu$, otherwise we obtain labyrinthine patterns that are essentially the same as those observed on fixed domains. 

Our second main conclusion, again holding for both models, is that given $\alpha<\nu$, a higher $\alpha$ leads to stripes that are parallel to the moving front. These stripes arise as a two-dimensional analogue of peak insertion between the moving front and the existing pattern. In contrast, a lower $\alpha$ permits the formation of stripes that are transverse to the moving boundary, either perpendicular or at a slanted angle. 
Along the ridges of the stripes, the extension of the stripes as the front moves forward can be seen as a two-dimensional analogue of a travelling wave.
For the Schnakenberg model, we were able to obtain an estimate for $\alpha_c$, the threshold for $\alpha$ that separates these two cases, by observing the structure of the stripe that evolves when the region competent to form a pattern is restricted in the $y$--direction, so that it 
%on a narrow, pseudo-two-dimensional domain that 
admits complicated patterns only in the $x$--direction. However, the separation between the two patterning regimes is not clear-cut, as we observe mixed behaviours for $\alpha \approx \alpha_c$. 
Moreover, despite its agreement in behaviour with the Schnakenberg model on broad two-dimensional domains, the CDIMA model does not display the peak splitting behaviour observed in the Schnakenberg model in one-dimensional and narrow two-dimensional domains.

\revision{Moreover, we found that in a model where both spot and stripe patterns are possible (e.g. Schnakenberg with $b=1.4$), spots appear mainly when the stripes bend and break apart. Since in cases B and C the wave of competency favours the formation of straighter, more aligned stripes, spots appear less often with these wave speeds.}

In future, it will be interesting to understand why the two models behave differently on one-dimensional domains, but nonetheless exhibit similar behaviours on non-narrow two-dimensional domains.
Another research direction, beyond the scope of this study, is to analytically determine the value of the critical threshold $\alpha_c$. We speculate that this could be done by approximating the dynamics of the two-dimensional system along a stripe as a modified one-dimensional system, and using a wavelet analysis to determine the existence of a travelling wave solution. 

We now compare the pattern selection mechanism investigated in this paper with other mechanisms proposed in earlier works.
In our model, we have chosen a wave of competency that advances in the $x$--direction only. However, the conclusions regarding alignment of the stripes with respect to the wave remain the same if the pattern subdomain expands isotropically (such as an expanding circle, see \ref{apx:circular_domain}), meaning that this mechanism does not rely on any anisotropy.
In contrast, for the anistropic diffusion mechanism considered by \citet{shoji2002DirectionalityStripesFormed}, where the two morphogens have potentially differing preferred directions of diffusion, the anisotropy is essential for alignment.
\revision{\citet{hiscock2015OrientationTuringlikePatterns} considered two mechanisms for pattern alignment, which were the existence of spatial gradients in the kinetic parameters, and anisotropies in diffusion or growth, which was also considered by \citet{krause2019InfluenceCurvatureGrowth}, along with the geometric curvature of the underlying domain.} These mechanisms were able to select between  disordered labyrinthine patterns and one particular alignment for the stripes, but not between multiple possible directions, as our mechanism is capable of doing.
Therefore, the mechanism we have proposed behaves differently compared to the previously proposed mechanisms for controlling the alignment and directionality of stripes. 

\revision{We would also like to compare the wave of competency mechanism with DDI models posed on growing domains. 
While the expanding competent subdomain behind a wave of competency might be conceptually similar to growing domain models, they have different biological motivations, are not mathematically equivalent, and differ in behaviours.
Models with isotropic growth, such as those studied in \citep{crampin1999ReactionDiffusionGrowing,vangorder2021TuringConditionsPattern,kulesa1996ModelMechanismSpatial,buttenschon2020CellSizeMechanical}, consider a scenario where the tissue underlying the patterns is itself expanding. This mechanism results in a dilution effect which acts on every point in the domain, causing patterns in the interior to continue to change as the domain grows. In contrast, a wave of competency only affects patterns near the boundary between the competent and incompetent subdomains, and interior patterns remain static once the wavefront has moved sufficiently far away.
In one spatial dimension, \citet{crampin1999ReactionDiffusionGrowing} showed that the multi-spike pattern in the Schnakenberg model with exponential domain growth undergoes frequency-doubling, where every spike splits into two once the domain has grown sufficiently large. This is in contrast to the behaviour behind a wave of competency, where only the leading spike can split. In two spatial dimensions, \citet{vangorder2021TuringConditionsPattern} showed that the Schnakenberg model with linear domain growth was able to produce disorganised, mixed spot-stripe patterns at low growth rate, and stripes parallel to the domain boundary at high growth rate. This behaviour is largely similar to our cases A and B. However, they did not observe stripes perpendicular to the domain boundary (our case C).
\citet{crampin2002PatternFormationReactiondiffusion} considered apical growth, which is a form of non-uniform growth that only happens near the domain boundary. Although apical growth bears some resemblance to the wave of competency, and produces similar behaviours in one dimension, it is not mathematically equivalent to the wave of competency model, and it remains to be seen whether these models have similar behaviours in two dimensions.
}

\revision{Further generalizations of the wave of competency model are possible.}
In this paper, the wave of competency has a simple shape,  and its movement is independent of the morphogen concentrations.
In a real biological system, the wave speed and the shape of the wave front may be coupled with the morphogen concentrations. For example, in \citep{liu2021SpotsStripesSpiral}, the patterning domain is interpreted as the interior of a cell, and the morphogens, through interactions with the actin cytoskeleton, deform the cell. 
It would be interesting to compare behaviours in this kind of dynamic domain under zero-flux boundary conditions, with behaviours in a domain bounded by a moving wave of competency.

In summary, we have demonstrated that modulation of the speed of propagation of a wave of competency can robustly select patterns in DDI models. 
These results have implications for patterning in a number of biological systems where a wave of competency has been either identified, or could exist.
Our results also pave the way for future studies that aim to control the directionality and alignment of DDI patterns by modulating the wave speed. More complex patterns may be obtained by also varying the shape of the wave front.

\section*{Acknowledgement}
YL is supported by the Natural Sciences and Engineering Research Council of Canada (NSERC) through the Postgraduate Scholarships – Doctoral program, reference number PGSD3-535584-2019.
REB acknowledges the Royal Society for a Wolfson Research Merit Award.

\appendix

\revision{
\section{Supplementary materials}
Animations corresponding to the solutions presented in all figures, as well as the Matlab code used to produce the figures and animations, are provided as supplementary materials available to be downloaded from the journal website.
The code and animations are also provided on Github at \url{github.com/liuyue002/turing_expanding_domain}.
}

\revision{
\section{Linear stability analysis}\label{apx:lsa}

In this appendix we perform a linear stability analysis of the Schnakenberg model. We use these results to guide parameter value selection.
The methods employed here follow \citet{turing1952ChemicalBasisMorphogenesis}, are entirely standard, and presented for completeness.

Firstly, we linearise the system of PDEs (Eq.~\eqref{sys:generic_rd}) around a spatially uniform steady state $(u_*, v_*)$ by letting $(U,V)=(u,v)-(u_*,v_*)$. The linearized system can be written as
\begin{equation}
\pd{}{t} \begin{bmatrix}U\\V\end{bmatrix} = D \pd{ ^2}{x^2}\begin{bmatrix}U\\V\end{bmatrix} + J \begin{bmatrix}U\\V\end{bmatrix},
\end{equation}
where $D=diag(D_u,D_v)$, and $J$ is the Jacobian of the system at $(u_*,v_*)$. For a DDI, we require that $(u_*, v_*)$ is stable in the absence of diffusion, that is, the eigenvalues of $J$ must have negative real parts, but is unstable in the presence of diffusion. To investigate linear stability, we impose an ansatz,
\begin{align*}
U &= a \exp[\sigma t + i (k_x x+ k_y y)], \\
V &= b \exp[\sigma t + i (k_x x+ k_y y)]. 
\end{align*}
The zero-flux boundary conditions imply that only modes with 
\[k_x = \frac{n^2 \pi^2}{L_x^2},\quad k_y = \frac{m^2 \pi^2}{L_y^2},\]
for some integers $m$, $n$ (where at least one of these is non-zero) are admissible.
The real part of $\sigma$ determines the linear stability of $(u_*, v_*)$, that is, if a non-trivial solution with $Re(\sigma)>0$ exists for any admissible $k_x, k_y$, then $(u_*, v_*)$ is linearly unstable, and a DDI can occur.

If we let $k^2=k_x^2 + k_y^2$, then it can be shown that $\sigma=\sigma(k^2)$ is the eigenvalue of $M=J-k^2 D$, and the condition for a DDI can be written as (for detailed derivation, see \citep[Ch.~11.4]{leah_book})
\begin{equation}
J_{11}D_v+J_{22}D_u > 2 \sqrt{D_u D_v \det(J)}, \label{eqn:turing_condition}
\end{equation}
where $J_{ij}$ denote the $(i,j)^{th}$ entry of the Jacobian $J$.

For the Schnakenberg model (Eq.~\eqref{sys:schnackenberg}), where the uniform steady state $(u_*, v_*)$ is given in Eq.~\eqref{eqn:schnackenberg_hss}, we find that the inequality \eqref{eqn:turing_condition} simplifies to 
\[\frac{2D_v b/u_* - D_u u_*^2 - D_v}{2\sqrt{D_v D_u}}-u_* > 0 ,\]
which holds for the parameter values given in Eq.~\eqref{eqn:schnackenberg_params} with $W=0$, but does not hold when $W=\Wmax$. This is the desired property for our choice of the parameters. Similar calculations can be done for the CDIMA model. 

Moreover, in the case where a DDI is possible, by plotting the real parts of the eigenvalues of $M$ as a function of $k$, we can determine the range of wave numbers $k$ for which the corresponding mode is unstable.
These are the only modes that will initially grow from the initial conditions.
For the Schnakenberg model with $b=1.4$, the range is $k \in (0.40, 0.82)$, and with $b=1.6$, the range is $k \in (0.16,0.24)$.

}

\section{Simulations on two-dimensional circular patterning subdomain}\label{apx:circular_domain}

In this appendix we simulate the CDIMA (Eq.~\eqref{sys:cdima}) and Schnakenberg models (Eq.~\eqref{sys:schnackenberg}) on a circular patterning subdomain that expands radially at constant speeds, that is,
\begin{equation}
\Omega'(t) = \{(x,y) | x^2 + y^2 \leq \rho(t)^2\},\quad \rho(t)=\min(\alpha t, L_x/2, L_y/2), \label{eqn:circular}
\end{equation}
instead of the rectangular patterning subdomain used in the main text. 
This choice matches the experimental and simulation set up of \citet{konow2019TuringPatternsRadially}, which enable us to compare our results.

\begin{figure}[H]
    \centering
    \includegraphics[width=0.9\textwidth]{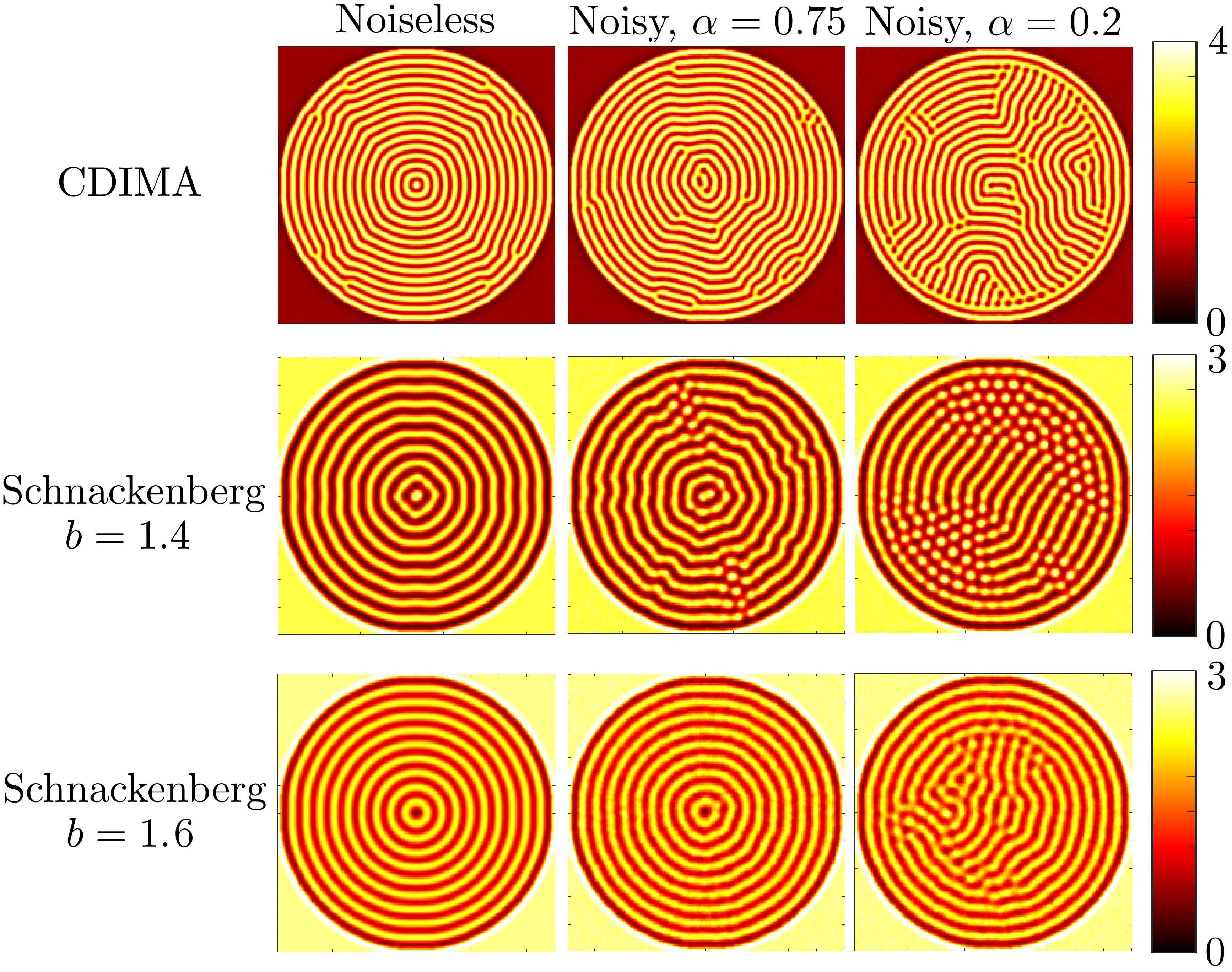}
    \caption{The steady state pattern produced with a circular patterning subdomain, as described in Eq.~\eqref{eqn:circular}. The system is initialized by perturbing the homogeneous steady state with noise. In the first column, without on-going noise, the final pattern takes the form of concentric rings regardless of the value of $\alpha$ (the plot shown is produced with $\alpha=0.75$). In the second and third columns, with the on-going noise we observe different qualitative behaviours for different values of $\alpha$.
    }
    \label{fig:2d_circular}
\end{figure}

The steady state patterns are shown in Fig.~\ref{fig:2d_circular}. For simulations without on-going noise, the resulting patterns are concentric rings regardless of the wave speed $\alpha$. With on-going noise, for the CDIMA model, we observe mostly concentric rings when the wave speed $\alpha$ is sufficiently fast, and mostly stripes transversal to the circular boundary when the wave speed is low. This is consistent with the observations from \citep{konow2019TuringPatternsRadially}.

The Schnakenberg model with $b=1.6$ exhibit mostly the same qualitative behaviour as the CDIMA model.
An interesting difference is that, with a low wave speed $\alpha$, despite producing a pattern similar to the CDIMA model at first, the stripes near the circular boundary eventually evolves to rings once the wave of competency stops, while the stripes in the center of the domain remain in a radial direction.
With $b=1.4$, the Schnakenberg model tends to produce more spots.
These are consistent with our observations in Section~\ref{sec:numerics} on a rectangular patterning subdomain, where the vertical stripes parallel to the linear wave front plays the same role as the rings in the circular setting. 

\section{Simulations with higher magnitude of noise}\label{apx:high_noise}

In this appendix we show the effect of higher levels of on-going noise on the pattern. We simulate the Schnakenberg model on a fixed domain (that is, $\Omega'=\Omega$), with $b=1.4$, and other parameter values from Eq.~\eqref{eqn:schnackenberg_params}. We initialize the system with horizontal stripes as described in Eq.~\ref{eqn:horizontal_stripes}. 

\begin{figure}[H]
    \centering
    \includegraphics[width=0.9\textwidth]{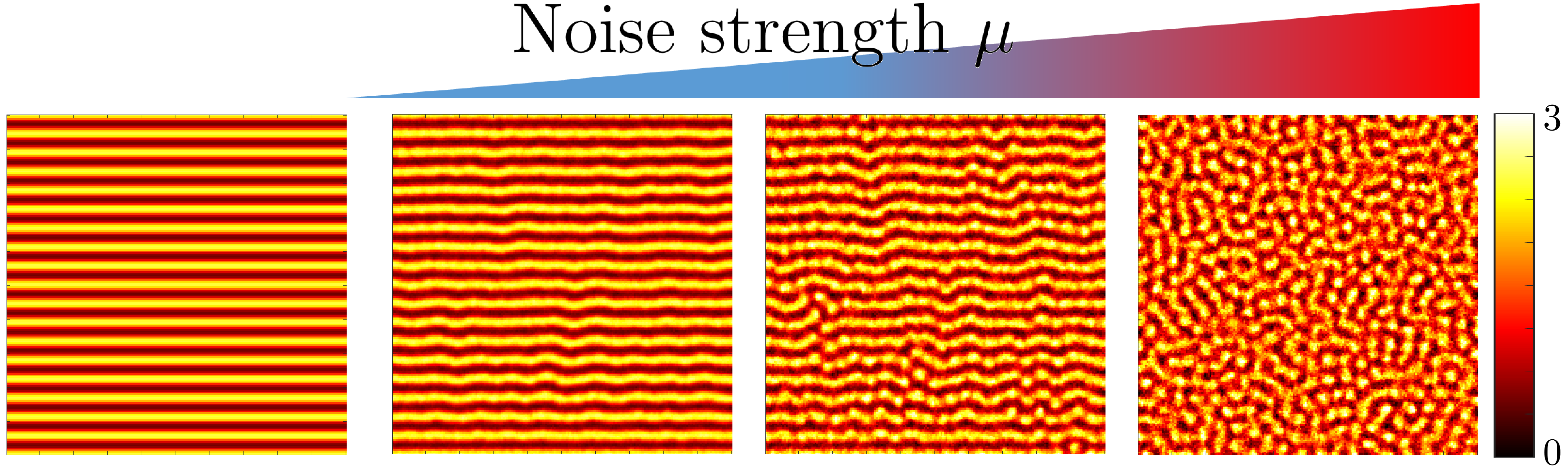}
    \caption{Snapshots of the simulations for the Schnakenberg model on a fixed domain, initialized at horizontal stripes, with $b=1.4$ and varying noise strength $\mu$, taken at $t=350$. From left to right: $\mu=0, 0.01, 0.03, 0.05$.}
    \label{fig:highnoise}
\end{figure}

\revision{
When the noise is sufficiently strong, the system does not settle to a steady state, so in Fig.~\ref{fig:highnoise} we chose to present snapshots of the system at $t=350$, which gives sufficient time for a steady state pattern to develop from perturbations to the homogeneous steady state in a fixed domain without on-going noise. 
Since on a two-dimensional domain, the system is multi-stable, over an infinitely long time scale even weak noise will eventually perturb the pattern into another steady state, disrupting the initial regular pattern. 
However, in biology cells have a finite time window in which to respond to morphogen signals, and model behaviour over an infinite time span is therefore not biologically relevant.
As such, we terminate the simulation on a time scale comparable to that of pattern formation from a perturbed uniform state to final steady state in a noiseless setting.}

We observe that with noise strength $\mu=0.01$, the initial horizontal pattern persists mostly undisturbed, only becoming slightly bent in certain places. With a stronger noise $\mu=0.03$, the horizontal stripes are still recognizable but significantly disrupted, and with $\mu=0.05$ the stripes are entirely unrecognizable, and the pattern evolves to a mess of spots and stripes that constantly change and shift as time continue to progress. We chose $\mu=0.01$ for the simulations in the main text so that the noise does not overwhelm the system.

\bibliographystyle{elsarticle-num-names} 
\bibliography{main}

\end{document}

%% file: my_include.tex
\usepackage{float}
\usepackage[usenames, dvipsnames]{xcolor}
\usepackage{graphicx}
\usepackage[multidot]{grffile}
\usepackage{subcaption}
\usepackage[final]{pdfpages}
\usepackage{amssymb,amsmath}
\usepackage{mathtools} %for dcases
\usepackage{listings} %for pretty print code
\usepackage{enumitem} %for specifying label style for enumerate
  
\usepackage[unicode=true,
  linktocpage,
  linkbordercolor={0.5 0.5 1},
  citebordercolor={0.5 1 0.5},
  colorlinks=true,
  linkcolor=blue]{hyperref}

\newcommand{\pd}[2]{\frac{\partial #1}{\partial #2}}

\newcommand{\lb}{\left(}
\newcommand{\rb}{\right)}

\newcommand{\Cos}[1]{\cos \lb #1 \rb}
\newcommand{\Wmax}{W_{\text{max}}}

\newcommand{\revision}[1]{#1}

\newcommand{\specialcell}[2][c]{%
  \begin{tabular}[#1]{@{}c@{}}#2\end{tabular}}
  
\floatstyle{ruled}
\newfloat{Program}{htbp}{lop} %for including code, usage: \begin{Program}...\end{Program}

\usepackage[margin=2.5cm]{geometry}

%% file: main.bbl
\begin{thebibliography}{47}
\expandafter\ifx\csname natexlab\endcsname\relax\def\natexlab#1{#1}\fi
\providecommand{\url}[1]{\texttt{#1}}
\providecommand{\href}[2]{#2}
\providecommand{\path}[1]{#1}
\providecommand{\DOIprefix}{doi:}
\providecommand{\ArXivprefix}{arXiv:}
\providecommand{\URLprefix}{URL: }
\providecommand{\Pubmedprefix}{pmid:}
\providecommand{\doi}[1]{\href{http://dx.doi.org/#1}{\path{#1}}}
\providecommand{\Pubmed}[1]{\href{pmid:#1}{\path{#1}}}
\providecommand{\bibinfo}[2]{#2}
\ifx\xfnm\relax \def\xfnm[#1]{\unskip,\space#1}\fi
%Type = Article
\bibitem[{Turing(1952)}]{turing1952ChemicalBasisMorphogenesis}
\bibinfo{author}{A.~M. Turing},
\newblock \bibinfo{title}{The chemical basis of morphogenesis},
\newblock \bibinfo{journal}{Philosophical Transactions of the Royal Society of
  London. Series B, Biological Sciences} \bibinfo{volume}{237}
  (\bibinfo{year}{1952}) \bibinfo{pages}{37--72}.
  \DOIprefix\doi{10.1098/rstb.1952.0012}.
%Type = Book
\bibitem[{Murray(2001)}]{murray2001mathematical}
\bibinfo{author}{J.~D. Murray}, \bibinfo{title}{{{Mathematical Biology II:
  Spatial Models and Biomedical Applications}}}, volume~\bibinfo{volume}{3},
  \bibinfo{publisher}{Springer New York}, \bibinfo{year}{2001}.
%Type = Article
\bibitem[{Kealy and Wollkind(2012)}]{kealy2012NonlinearStabilityAnalysis}
\bibinfo{author}{B.~J. Kealy}, \bibinfo{author}{D.~J. Wollkind},
\newblock \bibinfo{title}{A {{Nonlinear Stability Analysis}} of {{Vegetative
  Turing Pattern Formation}} for an {{Interaction}}\textendash{{Diffusion
  Plant-Surface Water Model System}} in an {{Arid Flat Environment}}},
\newblock \bibinfo{journal}{Bulletin of Mathematical Biology}
  \bibinfo{volume}{74} (\bibinfo{year}{2012}) \bibinfo{pages}{803--833}.
  \DOIprefix\doi{10.1007/s11538-011-9688-7}.
%Type = Article
\bibitem[{Cartwright(2002)}]{cartwright2002LabyrinthineTuringPattern}
\bibinfo{author}{J.~H.~E. Cartwright},
\newblock \bibinfo{title}{Labyrinthine {{Turing Pattern Formation}} in the
  {{Cerebral Cortex}}},
\newblock \bibinfo{journal}{Journal of Theoretical Biology}
  \bibinfo{volume}{217} (\bibinfo{year}{2002}) \bibinfo{pages}{97--103}.
  \DOIprefix\doi{10.1006/jtbi.2002.3012}.
%Type = Article
\bibitem[{Watanabe and Kondo(2015)}]{watanabe2015PigmentPatterningFish}
\bibinfo{author}{M.~Watanabe}, \bibinfo{author}{S.~Kondo},
\newblock \bibinfo{title}{Is pigment patterning in fish skin determined by the
  {{Turing}} mechanism?},
\newblock \bibinfo{journal}{Trends in Genetics} \bibinfo{volume}{31}
  (\bibinfo{year}{2015}) \bibinfo{pages}{88--96}.
  \DOIprefix\doi{10.1016/j.tig.2014.11.005}.
%Type = Article
\bibitem[{Krause et~al.(2021)Krause, Gaffney, Maini, and
  Klika}]{krause2021IntroductionRecentProgress}
\bibinfo{author}{A.~L. Krause}, \bibinfo{author}{E.~A. Gaffney},
  \bibinfo{author}{P.~K. Maini}, \bibinfo{author}{V.~Klika},
\newblock \bibinfo{title}{Introduction to `{{Recent}} progress and open
  frontiers in {{Turing}}'s theory of morphogenesis'},
\newblock \bibinfo{journal}{Philosophical Transactions of the Royal Society A:
  Mathematical, Physical and Engineering Sciences} \bibinfo{volume}{379}
  (\bibinfo{year}{2021}) \bibinfo{pages}{20200280}.
  \DOIprefix\doi{10.1098/rsta.2020.0280}.
%Type = Article
\bibitem[{Barrass et~al.(2006)Barrass, Crampin, and
  Maini}]{barrass2006ModeTransitionsModel}
\bibinfo{author}{I.~Barrass}, \bibinfo{author}{E.~J. Crampin},
  \bibinfo{author}{P.~K. Maini},
\newblock \bibinfo{title}{Mode transitions in a model reaction-diffusion system
  driven by domain growth and noise},
\newblock \bibinfo{journal}{Bulletin of Mathematical Biology}
  \bibinfo{volume}{68} (\bibinfo{year}{2006}) \bibinfo{pages}{981--995}.
  \DOIprefix\doi{10.1007/s11538-006-9106-8}.
%Type = Article
\bibitem[{Madzvamuse and
  Maini(2007)}]{madzvamuse2007VelocityinducedNumericalSolutions}
\bibinfo{author}{A.~Madzvamuse}, \bibinfo{author}{P.~K. Maini},
\newblock \bibinfo{title}{Velocity-induced numerical solutions of
  reaction-diffusion systems on continuously growing domains},
\newblock \bibinfo{journal}{Journal of Computational Physics}
  \bibinfo{volume}{225} (\bibinfo{year}{2007}) \bibinfo{pages}{100--119}.
  \DOIprefix\doi{10.1016/j.jcp.2006.11.022}.
%Type = Article
\bibitem[{Crampin et~al.(1999)Crampin, Gaffney, and
  Maini}]{crampin1999ReactionDiffusionGrowing}
\bibinfo{author}{E.~J. Crampin}, \bibinfo{author}{E.~A. Gaffney},
  \bibinfo{author}{P.~K. Maini},
\newblock \bibinfo{title}{Reaction and diffusion on growing domains: scenarios
  for robust pattern formation},
\newblock \bibinfo{journal}{Bulletin of Mathematical Biology}
  \bibinfo{volume}{61} (\bibinfo{year}{1999}) \bibinfo{pages}{1093--1120}.
  \DOIprefix\doi{10.1006/bulm.1999.0131}.
%Type = Article
\bibitem[{Crampin et~al.(2002)Crampin, Hackborn, and
  Maini}]{crampin2002PatternFormationReactiondiffusion}
\bibinfo{author}{E.~J. Crampin}, \bibinfo{author}{W.~W. Hackborn},
  \bibinfo{author}{P.~K. Maini},
\newblock \bibinfo{title}{Pattern formation in reaction-diffusion models with
  nonuniform domain growth},
\newblock \bibinfo{journal}{Bulletin of Mathematical Biology}
  \bibinfo{volume}{64} (\bibinfo{year}{2002}) \bibinfo{pages}{747--769}.
  \DOIprefix\doi{10.1006/bulm.2002.0295}.
%Type = Article
\bibitem[{Van~Gorder et~al.(2021)Van~Gorder, Klika, and
  Krause}]{vangorder2021TuringConditionsPattern}
\bibinfo{author}{R.~A. Van~Gorder}, \bibinfo{author}{V.~Klika},
  \bibinfo{author}{A.~L. Krause},
\newblock \bibinfo{title}{Turing conditions for pattern forming systems on
  evolving manifolds},
\newblock \bibinfo{journal}{Journal of Mathematical Biology}
  \bibinfo{volume}{82} (\bibinfo{year}{2021}) \bibinfo{pages}{4}.
  \DOIprefix\doi{10.1007/s00285-021-01552-y}.
%Type = Article
\bibitem[{Bailleul et~al.(2019)Bailleul, Curantz, Dinh, Hidalgo, Touboul, and
  Manceau}]{bailleul2019SymmetryBreakingEmbryonic}
\bibinfo{author}{R.~Bailleul}, \bibinfo{author}{C.~Curantz},
  \bibinfo{author}{C.~D.-T. Dinh}, \bibinfo{author}{M.~Hidalgo},
  \bibinfo{author}{J.~Touboul}, \bibinfo{author}{M.~Manceau},
\newblock \bibinfo{title}{Symmetry breaking in the embryonic skin triggers
  directional and sequential plumage patterning},
\newblock \bibinfo{journal}{PLOS Biology} \bibinfo{volume}{17}
  (\bibinfo{year}{2019}) \bibinfo{pages}{e3000448}.
  \DOIprefix\doi{10.1371/journal.pbio.3000448}.
%Type = Article
\bibitem[{Jung et~al.(1998)Jung, {Francis-West}, Widelitz, Jiang,
  {Ting-Berreth}, Tickle, Wolpert, and Chuong}]{jung1998LocalInhibitoryAction}
\bibinfo{author}{H.-S. Jung}, \bibinfo{author}{P.~H. {Francis-West}},
  \bibinfo{author}{R.~B. Widelitz}, \bibinfo{author}{T.-X. Jiang},
  \bibinfo{author}{S.~{Ting-Berreth}}, \bibinfo{author}{C.~Tickle},
  \bibinfo{author}{L.~Wolpert}, \bibinfo{author}{C.-M. Chuong},
\newblock \bibinfo{title}{Local inhibitory action of {{BMPs}} and their
  relationships with activators in feather formation: implications for periodic
  patterning},
\newblock \bibinfo{journal}{Developmental Biology} \bibinfo{volume}{196}
  (\bibinfo{year}{1998}) \bibinfo{pages}{11--23}.
  \DOIprefix\doi{10.1006/dbio.1998.8850}.
%Type = Article
\bibitem[{Jiang et~al.(1999)Jiang, Jung, Widelitz, and
  Chuong}]{jiang1999SelforganizationPeriodicPatterns}
\bibinfo{author}{T.~Jiang}, \bibinfo{author}{H.~Jung},
  \bibinfo{author}{R.~Widelitz}, \bibinfo{author}{C.~Chuong},
\newblock \bibinfo{title}{Self-organization of periodic patterns by dissociated
  feather mesenchymal cells and the regulation of size, number and spacing of
  primordia},
\newblock \bibinfo{journal}{Development} \bibinfo{volume}{126}
  (\bibinfo{year}{1999}) \bibinfo{pages}{4997--5009}.
  \DOIprefix\doi{10.1242/dev.126.22.4997}.
%Type = Article
\bibitem[{Mou et~al.(2011)Mou, Pitel, Gourichon, Vignoles, Tzika, Tato, Yu,
  Burt, Bed'hom, {Tixier-Boichard}, Painter, and
  Headon}]{mou2011CrypticPatterningAvian}
\bibinfo{author}{C.~Mou}, \bibinfo{author}{F.~Pitel},
  \bibinfo{author}{D.~Gourichon}, \bibinfo{author}{F.~Vignoles},
  \bibinfo{author}{A.~Tzika}, \bibinfo{author}{P.~Tato},
  \bibinfo{author}{L.~Yu}, \bibinfo{author}{D.~W. Burt},
  \bibinfo{author}{B.~Bed'hom}, \bibinfo{author}{M.~{Tixier-Boichard}},
  \bibinfo{author}{K.~J. Painter}, \bibinfo{author}{D.~J. Headon},
\newblock \bibinfo{title}{Cryptic patterning of avian skin confers a
  developmental facility for loss of neck feathering},
\newblock \bibinfo{journal}{PLOS Biology} \bibinfo{volume}{9}
  (\bibinfo{year}{2011}) \bibinfo{pages}{e1001028}.
  \DOIprefix\doi{10.1371/journal.pbio.1001028}.
%Type = Article
\bibitem[{Kondo and Asai(1995)}]{kondo1995ReactionDiffusionWave}
\bibinfo{author}{S.~Kondo}, \bibinfo{author}{R.~Asai},
\newblock \bibinfo{title}{A reaction-diffusion wave on the skin of the marine
  angelfish {{Pomacanthus}}},
\newblock \bibinfo{journal}{Nature} \bibinfo{volume}{376}
  (\bibinfo{year}{1995}) \bibinfo{pages}{765--768}.
  \DOIprefix\doi{10.1038/376765a0}.
%Type = Article
\bibitem[{Owen et~al.(2020)Owen, Kelsh, and
  Yates}]{owen2020QuantitativeModellingApproach}
\bibinfo{author}{J.~P. Owen}, \bibinfo{author}{R.~N. Kelsh},
  \bibinfo{author}{C.~A. Yates},
\newblock \bibinfo{title}{A quantitative modelling approach to zebrafish
  pigment pattern formation},
\newblock \bibinfo{journal}{eLife} \bibinfo{volume}{9} (\bibinfo{year}{2020})
  \bibinfo{pages}{e52998}. \DOIprefix\doi{10.7554/eLife.52998}.
%Type = Article
\bibitem[{Volkening and
  Sandstede(2015)}]{volkening2015ModellingStripeFormation}
\bibinfo{author}{A.~Volkening}, \bibinfo{author}{B.~Sandstede},
\newblock \bibinfo{title}{Modelling stripe formation in zebrafish: An
  agent-based approach},
\newblock \bibinfo{journal}{Journal of The Royal Society Interface}
  \bibinfo{volume}{12} (\bibinfo{year}{2015}) \bibinfo{pages}{20150812}.
  \DOIprefix\doi{10.1098/rsif.2015.0812}.
%Type = Article
\bibitem[{Frohnh{\"o}fer et~al.(2013)Frohnh{\"o}fer, Krauss, Maischein, and
  {N{\"u}sslein-Volhard}}]{frohnhofer2013IridophoresTheirInteractions}
\bibinfo{author}{H.~G. Frohnh{\"o}fer}, \bibinfo{author}{J.~Krauss},
  \bibinfo{author}{H.-M. Maischein},
  \bibinfo{author}{C.~{N{\"u}sslein-Volhard}},
\newblock \bibinfo{title}{Iridophores and their interactions with other
  chromatophores are required for stripe formation in zebrafish},
\newblock \bibinfo{journal}{Development} \bibinfo{volume}{140}
  (\bibinfo{year}{2013}) \bibinfo{pages}{2997--3007}.
  \DOIprefix\doi{10.1242/dev.096719}.
%Type = Misc
\bibitem[{Azul(2005)}]{zebrafishpic}
\bibinfo{author}{Azul}, \bibinfo{title}{Zebrafish}, \bibinfo{year}{2005}.
  \URLprefix \url{https://commons.wikimedia.org/wiki/File:Zebrafisch.jpg},
  \bibinfo{note}{image under CC-0 license.}
%Type = Misc
\bibitem[{Zerpe(2019)}]{trevallypic}
\bibinfo{author}{R.~Zerpe}, \bibinfo{title}{Golden trevally (gnathanodon
  speciosus)}, \bibinfo{year}{2019}. \URLprefix
  \url{https://www.flickr.com/photos/krokodiver/47728328672/},
  \bibinfo{note}{image under CC-BY license.}
%Type = Article
\bibitem[{Barrio et~al.(2009)Barrio, Baker, Vaughan, Tribuzy, {de Carvalho},
  Bassanezi, and Maini}]{barrio2009ModelingSkinPattern}
\bibinfo{author}{R.~A. Barrio}, \bibinfo{author}{R.~E. Baker},
  \bibinfo{author}{B.~Vaughan}, \bibinfo{author}{K.~Tribuzy},
  \bibinfo{author}{M.~R. {de Carvalho}}, \bibinfo{author}{R.~Bassanezi},
  \bibinfo{author}{P.~K. Maini},
\newblock \bibinfo{title}{Modeling the skin pattern of fishes},
\newblock \bibinfo{journal}{Physical Review E} \bibinfo{volume}{79}
  (\bibinfo{year}{2009}) \bibinfo{pages}{031908}.
  \DOIprefix\doi{10.1103/PhysRevE.79.031908}.
%Type = Article
\bibitem[{Shoji et~al.(2002)Shoji, Iwasa, Mochizuki, and
  Kondo}]{shoji2002DirectionalityStripesFormed}
\bibinfo{author}{H.~Shoji}, \bibinfo{author}{Y.~Iwasa},
  \bibinfo{author}{A.~Mochizuki}, \bibinfo{author}{S.~Kondo},
\newblock \bibinfo{title}{Directionality of stripes formed by anisotropic
  reaction-diffusion models},
\newblock \bibinfo{journal}{Journal of Theoretical Biology}
  \bibinfo{volume}{214} (\bibinfo{year}{2002}) \bibinfo{pages}{549--561}.
  \DOIprefix\doi{10.1006/jtbi.2001.2480}.
%Type = Article
\bibitem[{Hiscock and Megason(2015)}]{hiscock2015OrientationTuringlikePatterns}
\bibinfo{author}{T.~W. Hiscock}, \bibinfo{author}{S.~G. Megason},
\newblock \bibinfo{title}{Orientation of {{Turing}}-like patterns by morphogen
  gradients and tissue anisotropies},
\newblock \bibinfo{journal}{Cell Systems} \bibinfo{volume}{1}
  (\bibinfo{year}{2015}) \bibinfo{pages}{408--416}.
  \DOIprefix\doi{10.1016/j.cels.2015.12.001}.
%Type = Article
\bibitem[{Nakamasu et~al.(2009)Nakamasu, Takahashi, Kanbe, and
  Kondo}]{nakamasu2009InteractionsZebrafishPigment}
\bibinfo{author}{A.~Nakamasu}, \bibinfo{author}{G.~Takahashi},
  \bibinfo{author}{A.~Kanbe}, \bibinfo{author}{S.~Kondo},
\newblock \bibinfo{title}{Interactions between zebrafish pigment cells
  responsible for the generation of {{Turing}} patterns},
\newblock \bibinfo{journal}{Proceedings of the National Academy of Sciences}
  \bibinfo{volume}{106} (\bibinfo{year}{2009}) \bibinfo{pages}{8429--8434}.
  \DOIprefix\doi{10.1073/pnas.0808622106}.
%Type = Article
\bibitem[{Page et~al.(2005)Page, Maini, and
  Monk}]{page2005ComplexPatternFormation}
\bibinfo{author}{K.~M. Page}, \bibinfo{author}{P.~K. Maini},
  \bibinfo{author}{N.~A.~M. Monk},
\newblock \bibinfo{title}{Complex pattern formation in reaction-diffusion
  systems with spatially varying parameters},
\newblock \bibinfo{journal}{Physica D: Nonlinear Phenomena}
  \bibinfo{volume}{202} (\bibinfo{year}{2005}) \bibinfo{pages}{95--115}.
  \DOIprefix\doi{10.1016/j.physd.2005.01.022}.
%Type = Article
\bibitem[{Konow et~al.(2019)Konow, H.~Somberg, Chavez, R.~Epstein, and
  Dolnik}]{konow2019TuringPatternsRadially}
\bibinfo{author}{C.~Konow}, \bibinfo{author}{N.~H.~Somberg},
  \bibinfo{author}{J.~Chavez}, \bibinfo{author}{I.~R.~Epstein},
  \bibinfo{author}{M.~Dolnik},
\newblock \bibinfo{title}{Turing patterns on radially growing domains:
  Experiments and simulations},
\newblock \bibinfo{journal}{Physical Chemistry Chemical Physics}
  \bibinfo{volume}{21} (\bibinfo{year}{2019}) \bibinfo{pages}{6718--6724}.
  \DOIprefix\doi{10.1039/C8CP07797E}.
%Type = Article
\bibitem[{Schnakenberg(1979)}]{schnakenberg1979SimpleChemicalReaction}
\bibinfo{author}{J.~Schnakenberg},
\newblock \bibinfo{title}{Simple chemical reaction systems with limit cycle
  behaviour},
\newblock \bibinfo{journal}{Journal of Theoretical Biology}
  \bibinfo{volume}{81} (\bibinfo{year}{1979}) \bibinfo{pages}{389--400}.
  \DOIprefix\doi{10.1016/0022-5193(79)90042-0}.
%Type = Article
\bibitem[{Barrio(1999)}]{barrio1999TwodimensionalNumericalStudy}
\bibinfo{author}{R.~Barrio},
\newblock \bibinfo{title}{A two-dimensional numerical study of spatial pattern
  formation in interacting turing systems},
\newblock \bibinfo{journal}{Bulletin of Mathematical Biology}
  \bibinfo{volume}{61} (\bibinfo{year}{1999}) \bibinfo{pages}{483--505}.
  \DOIprefix\doi{10.1006/bulm.1998.0093}.
%Type = Article
\bibitem[{Varea et~al.(1997)Varea, Arag{\'o}n, and
  Barrio}]{varea1997ConfinedTuringPatterns}
\bibinfo{author}{C.~Varea}, \bibinfo{author}{J.~L. Arag{\'o}n},
  \bibinfo{author}{R.~A. Barrio},
\newblock \bibinfo{title}{Confined {{Turing}} patterns in growing systems},
\newblock \bibinfo{journal}{Physical Review E} \bibinfo{volume}{56}
  (\bibinfo{year}{1997}) \bibinfo{pages}{1250--1253}.
  \DOIprefix\doi{10.1103/PhysRevE.56.1250}.
%Type = Article
\bibitem[{Bunow et~al.(1980)Bunow, Kernevez, Joly, and
  Thomas}]{bunow1980PatternFormationReactiondiffusion}
\bibinfo{author}{B.~Bunow}, \bibinfo{author}{J.-P. Kernevez},
  \bibinfo{author}{G.~Joly}, \bibinfo{author}{D.~Thomas},
\newblock \bibinfo{title}{Pattern formation by reaction-diffusion
  instabilities: application to morphogenesis in {{Drosophila}}},
\newblock \bibinfo{journal}{Journal of Theoretical Biology}
  \bibinfo{volume}{84} (\bibinfo{year}{1980}) \bibinfo{pages}{629--649}.
  \DOIprefix\doi{10.1016/S0022-5193(80)80024-5}.
%Type = Article
\bibitem[{Gierer and Meinhardt(1972)}]{gierer1972TheoryBiologicalPattern}
\bibinfo{author}{A.~Gierer}, \bibinfo{author}{H.~Meinhardt},
\newblock \bibinfo{title}{A theory of biological pattern formation},
\newblock \bibinfo{journal}{Kybernetik} \bibinfo{volume}{12}
  (\bibinfo{year}{1972}) \bibinfo{pages}{30--39}.
  \DOIprefix\doi{10.1007/BF00289234}.
%Type = Article
\bibitem[{Iron et~al.(2004)Iron, Wei, and
  Winter}]{iron2004StabilityAnalysisTuring}
\bibinfo{author}{D.~Iron}, \bibinfo{author}{J.~Wei},
  \bibinfo{author}{M.~Winter},
\newblock \bibinfo{title}{Stability analysis of {{Turing}} patterns generated
  by the {{Schnakenberg}} model},
\newblock \bibinfo{journal}{Journal of Mathematical Biology}
  \bibinfo{volume}{49} (\bibinfo{year}{2004}) \bibinfo{pages}{358--390}.
  \DOIprefix\doi{10.1007/s00285-003-0258-y}.
%Type = Article
\bibitem[{Ermentrout and Enderby(1991)}]{ermentrout1991StripesSpotsNonlineara}
\bibinfo{author}{B.~Ermentrout}, \bibinfo{author}{J.~E. Enderby},
\newblock \bibinfo{title}{Stripes or spots? {{Nonlinear}} effects in
  bifurcation of reaction-diffusion equations on the square},
\newblock \bibinfo{journal}{Proceedings of the Royal Society of London. Series
  A: Mathematical and Physical Sciences} \bibinfo{volume}{434}
  (\bibinfo{year}{1991}) \bibinfo{pages}{413--417}.
  \DOIprefix\doi{10.1098/rspa.1991.0100}.
%Type = Article
\bibitem[{Nagorcka and Mooney(1992)}]{nagorcka1992StripesSpotsPrepatterns}
\bibinfo{author}{B.~N. Nagorcka}, \bibinfo{author}{J.~R. Mooney},
\newblock \bibinfo{title}{From stripes to spots: {{Prepatterns}} which can be
  produced in the skin by a reaction-diffusion system},
\newblock \bibinfo{journal}{Mathematical Medicine and Biology: A Journal of the
  IMA} \bibinfo{volume}{9} (\bibinfo{year}{1992}) \bibinfo{pages}{249--267}.
  \DOIprefix\doi{10.1093/imammb/9.4.249}.
%Type = Article
\bibitem[{Lengyel and Epstein(1991)}]{lengyel1991ModelingTuringStructures}
\bibinfo{author}{I.~Lengyel}, \bibinfo{author}{I.~R. Epstein},
\newblock \bibinfo{title}{Modeling of {{Turing}} structures in the
  {{Chlorite}}\textendash{{Iodide}}\textendash{{Malonic
  Acid}}\textendash{{Starch reaction system}}},
\newblock \bibinfo{journal}{Science} \bibinfo{volume}{251}
  (\bibinfo{year}{1991}) \bibinfo{pages}{650--652}.
  \DOIprefix\doi{10.1126/science.251.4994.650}.
%Type = Article
\bibitem[{Bard and Lauder(1974)}]{bard1974HowWellDoes}
\bibinfo{author}{J.~Bard}, \bibinfo{author}{I.~Lauder},
\newblock \bibinfo{title}{How well does {{Turing}}'s theory of morphogenesis
  work?},
\newblock \bibinfo{journal}{Journal of Theoretical Biology}
  \bibinfo{volume}{45} (\bibinfo{year}{1974}) \bibinfo{pages}{501--531}.
  \DOIprefix\doi{10.1016/0022-5193(74)90128-3}.
%Type = Article
\bibitem[{Ruuth(1995)}]{ruuth1995ImplicitexplicitMethodsReactiondiffusion}
\bibinfo{author}{S.~J. Ruuth},
\newblock \bibinfo{title}{Implicit-explicit methods for reaction-diffusion
  problems in pattern formation},
\newblock \bibinfo{journal}{Journal of Mathematical Biology}
  \bibinfo{volume}{34} (\bibinfo{year}{1995}) \bibinfo{pages}{148--176}.
  \DOIprefix\doi{10.1007/BF00178771}.
%Type = Article
\bibitem[{Ascher et~al.(1995)Ascher, Ruuth, and
  Wetton}]{ascher1995ImplicitExplicitMethodsTimeDependent}
\bibinfo{author}{U.~M. Ascher}, \bibinfo{author}{S.~J. Ruuth},
  \bibinfo{author}{B.~T.~R. Wetton},
\newblock \bibinfo{title}{Implicit-explicit methods for time-dependent partial
  differential equations},
\newblock \bibinfo{journal}{SIAM Journal on Numerical Analysis}
  \bibinfo{volume}{32} (\bibinfo{year}{1995}) \bibinfo{pages}{797--823}.
  \DOIprefix\doi{10.1137/0732037}.
%Type = Book
\bibitem[{Canuto et~al.(1988)Canuto, Hussaini, Quarteroni, and
  Zang}]{canuto1988SpectralMethodsFluid}
\bibinfo{author}{C.~Canuto}, \bibinfo{author}{M.~Y. Hussaini},
  \bibinfo{author}{A.~Quarteroni}, \bibinfo{author}{T.~A. Zang},
  \bibinfo{title}{Spectral {{Methods}} in {{Fluid Dynamics}}},
  \bibinfo{publisher}{{Springer Berlin Heidelberg}}, \bibinfo{address}{{Berlin,
  Heidelberg}}, \bibinfo{year}{1988}.
  \DOIprefix\doi{10.1007/978-3-642-84108-8}.
%Type = Article
\bibitem[{Kamrani and
  Mohammad~Hosseini(2010)}]{kamrani2010RoleCoefficientsGeneral}
\bibinfo{author}{M.~Kamrani}, \bibinfo{author}{S.~Mohammad~Hosseini},
\newblock \bibinfo{title}{The role of coefficients of a general {{SPDE}} on the
  stability and convergence of a finite difference method},
\newblock \bibinfo{journal}{Journal of Computational and Applied Mathematics}
  \bibinfo{volume}{234} (\bibinfo{year}{2010}) \bibinfo{pages}{1426--1434}.
  \DOIprefix\doi{10.1016/j.cam.2010.02.018}.
%Type = Article
\bibitem[{Zouraris(2018)}]{zouraris2018CrankNicolsonFinite}
\bibinfo{author}{G.~E. Zouraris},
\newblock \bibinfo{title}{Crank--{{Nicolson}} finite element approximations for
  a linear stochastic fourth order equation with additive space-time white
  noise},
\newblock \bibinfo{journal}{SIAM Journal on Numerical Analysis}
  \bibinfo{volume}{56} (\bibinfo{year}{2018}) \bibinfo{pages}{838--858}.
  \DOIprefix\doi{10.1137/15M1053098}.
%Type = Article
\bibitem[{Krause et~al.(2019)Krause, Ellis, and
  Van~Gorder}]{krause2019InfluenceCurvatureGrowth}
\bibinfo{author}{A.~L. Krause}, \bibinfo{author}{M.~A. Ellis},
  \bibinfo{author}{R.~A. Van~Gorder},
\newblock \bibinfo{title}{Influence of curvature, growth, and anisotropy on the
  evolution of {{Turing}} patterns on growing manifolds},
\newblock \bibinfo{journal}{Bulletin of Mathematical Biology}
  \bibinfo{volume}{81} (\bibinfo{year}{2019}) \bibinfo{pages}{759--799}.
  \DOIprefix\doi{10.1007/s11538-018-0535-y}.
%Type = Article
\bibitem[{Kulesa et~al.(1996)Kulesa, Cruywagen, Lubkin, Main, Sneyd, Ferguson,
  and Murray}]{kulesa1996ModelMechanismSpatial}
\bibinfo{author}{P.~M. Kulesa}, \bibinfo{author}{G.~C. Cruywagen},
  \bibinfo{author}{S.~R. Lubkin}, \bibinfo{author}{P.~K. Main},
  \bibinfo{author}{J.~Sneyd}, \bibinfo{author}{M.~W.~J. Ferguson},
  \bibinfo{author}{J.~D. Murray},
\newblock \bibinfo{title}{On a model mechanism for the spatial patterning of
  teeth primordia in the {{Alligator}}},
\newblock \bibinfo{journal}{Journal of Theoretical Biology}
  \bibinfo{volume}{180} (\bibinfo{year}{1996}) \bibinfo{pages}{287--296}.
  \DOIprefix\doi{10.1006/jtbi.1996.0103}.
%Type = Article
\bibitem[{Buttensch{\"o}n et~al.(2020)Buttensch{\"o}n, Liu, and
  {Edelstein-Keshet}}]{buttenschon2020CellSizeMechanical}
\bibinfo{author}{A.~Buttensch{\"o}n}, \bibinfo{author}{Y.~Liu},
  \bibinfo{author}{L.~{Edelstein-Keshet}},
\newblock \bibinfo{title}{Cell size, mechanical tension, and {{GTPase}}
  signaling in the single cell},
\newblock \bibinfo{journal}{Bulletin of Mathematical Biology}
  \bibinfo{volume}{82} (\bibinfo{year}{2020}) \bibinfo{pages}{28}.
  \DOIprefix\doi{10.1007/s11538-020-00702-5}.
%Type = Article
\bibitem[{Liu et~al.(2021)Liu, Rens, and
  {Edelstein-Keshet}}]{liu2021SpotsStripesSpiral}
\bibinfo{author}{Y.~Liu}, \bibinfo{author}{E.~G. Rens},
  \bibinfo{author}{L.~{Edelstein-Keshet}},
\newblock \bibinfo{title}{Spots, stripes, and spiral waves in models for static
  and motile cells},
\newblock \bibinfo{journal}{Journal of Mathematical Biology}
  \bibinfo{volume}{82} (\bibinfo{year}{2021}) \bibinfo{pages}{28}.
  \DOIprefix\doi{10.1007/s00285-021-01550-0}.
%Type = Book
\bibitem[{{Edelstein-Keshet}(2005)}]{leah_book}
\bibinfo{author}{L.~{Edelstein-Keshet}}, \bibinfo{title}{Mathematical
  {{Models}} in {{Biology}}}, Classics in {{Applied Mathematics}},
  \bibinfo{publisher}{{Society for Industrial and Applied Mathematics}},
  \bibinfo{year}{2005}. \DOIprefix\doi{10.1137/1.9780898719147}.

\end{thebibliography}
